\documentclass[aps,preprint,nofootinbib,tightenlines,superscriptaddress]{revtex4}
\usepackage{epsfig}
\usepackage{amsfonts}
\usepackage{amsmath}
\usepackage{graphicx}
\newcommand{\beq}{\begin{equation}}
\newcommand{\eeq}{\end{equation}}
\newcommand{\bqa}{\begin{eqnarray}}
\newcommand{\eqa}{\end{eqnarray}}

\def\bfsigma{\mbox{\boldmath $\sigma$}}

\def\nnbe{\nonumber\\&=&}
\def\bold{\mathrm}
\begin{document}

\title{Radiative Leptonic $B_c\to \gamma \ell\bar\nu$ Decay in Effective Field Theory beyond  Leading Order}
\author{Wei Wang}
\author{Rui-Lin Zhu~\footnote{Corresponding author}}
\affiliation{
 INPAC, Shanghai Key Laboratory for Particle Physics and Cosmology, Department of Physics and Astronomy, Shanghai Jiao Tong University, Shanghai, 200240,   China}
 \affiliation{
State Key Laboratory of Theoretical Physics, Institute of Theoretical Physics, Chinese Academy of Sciences, Beijing 100190, China}

\email{wei.wang@sjtu.edu.cn}
\email{rlzhu@sjtu.edu.cn}

\begin{abstract}
We study the radiative leptonic $B_c\to \gamma\ell\bar\nu$ decays in the nonrelativistic QCD
effective field theory, and consider a  fast-moving photon. As a result the interactions with the heavy
quarks can  be  integrated out,  and thus we arrive at a factorization formula for the decay amplitude.
We calculate not only the relevant  short-distance coefficients at leading order and next-to-leading order
in $\alpha_s$, but also the nonrelativistic corrections at the order  $|\bold{v}|^2$ in our analysis.
We find out that the QCD corrections can sizably  decrease the branching ratio and thus  is of great
importance in extracting the long-distance operator matrix elements of $B_c$. For the phenomenological
application, we   present our results for  the photon energy, lepton energy and lepton-neutrino invariant
mass distribution.

\end{abstract}

\maketitle

\section{Introduction}

The search for new degrees of freedom can proceed  under two distinctive directions.
At the high energy frontier,  new particles have different signatures  with the standard
model (SM) particles, and  measurements of their production may provide definitive evidence
on their existence. On the other hand,  it is likely that   low energy
processes  will be influenced through  loop effects. Rare  decays of heavy mesons, with tiny
decay rates in the SM, are    sensitive to the new
degrees of freedom and thus can be exploited as indirect searches
of these unknown effects, for a recent review see Ref.~\cite{Wang:2014sba}.

The $B_c$ meson is the  unique pseudo-scalar meson that is long lived  and composed of two
different heavy flavors. Since this hadron is stable against strong interactions, its weak
decays provide a rich phenomena for the study of CKM matrix elements, and also a platform
to study the effects of weak interactions in a heavy quarkonium system~\cite{Brambilla:2004wf,Brambilla:2010cs}.
In the past decades  it has received  growing attentions since the first observation
by the CDF collaboration~\cite{Abe:1998wi}.  This can be particularly witnessed by the recent
LHCb measurements of the $B_c$ lifetime~\cite{Aaij:2014gka,Aaij:2014bva},
the decay widths of $B_c\to J/\psi\pi$ and $B_c\to J/\psi\ell\bar\nu$~\cite{Aaij:2014jxa,Aaij:2014ija},
and various other decay modes~\cite{Aaij:2014bla,Aaij:2014asa,Aaij:2013vcx,Aaij:2013oya}.
One may  expect that more decay channels of $B_c$ can  be measured by the LHCb,  ATLAS and CMS experiments
\cite{Bediaga:2012py,ATLAS:2012bja,Khachatryan:2014nfa}.

On theoretical side, various   approaches have been applied to calculate  the decay width of $B_c$ decays~~\cite{Du:1988ws,Colangelo:1992cx,Kiselev:1993ea,Choudhury:1998hs,Kiselev:1999sc,Nobes:2000pm,Ivanov:2000aj,Kiselev:2002vz,Ebert:2003cn,
Ebert:2003wc,Sun:2008ew,Zuo:2006re,Ivanov:2005fd,Wang:2007sxa,Wang:2007fs,Aliev:2006vs,Hernandez:2006gt,Huang:2007kb,Dhir:2008zz,Verma:2001hb,
Dhir:2008hh,Wang:2008xt,Wang:2009mi,Zhong:2014fma,Qiao:2012vt,Wen-Fei:2013uea,Rui:2014tpa,Shen:2014msa,Wu:2002ig,Huang:2008zg,Qiao:2012hp,Qiao:2011yz,Liu:2009qa,Liu:2011zzc,Xiao:2011zz,Xiao:2013lia,Wang:2012vna}, but most of them are   phenomenological.
Since  both constituents of the $B_c$ are heavy and  can only be treated nonrelativistically, an effective field theory can be established~~\cite{Bodwin:1994jh}.  Taking the $B_c\to J/\psi \ell\bar\nu$  as the example,
one may derive the conjectured non-relativistic QCD (NRQCD) factorization formula for its decay amplitude:
\begin{eqnarray}
{\cal A}(B_c\to J/\psi ) &\propto&  C_{ij} \langle 0|{\cal O}_{i}^{\bar cb}|\overline B_c\rangle \times  \langle J/\psi|{\cal O}_{j}^{\bar cc}|0\rangle,
\end{eqnarray}
where the   ${\cal O}_{i,j}^{ff'}$ are constructed by low energy operators. The short-distance, or {\it hard},  contributions at the length scale $1/m_{b,c}$  are encapsulated  into the coefficients $C_{ij}$ that can be computed  in perturbation theory.

The long-distance, or {\it soft} part of,  matrix elements  have to be extracted in  a nonperturbative approach, for instance the Lattice QCD simulation,  or constrained  by much simpler processes for instance the annihilation modes $B_c\to \ell\bar\nu$ and $B_c\to \gamma \ell\bar\nu$. However, the usefulness of the $B_c\to \ell\bar\nu$ is challenged by two aspects. Firstly its  decay rate  is given by
\begin{eqnarray}
 \Gamma(B_c\to \ell\bar\nu_\ell)&=& \frac{G_F^2}{8\pi} |V_{cb}|^2 f_{B_c}^2 m_{B_c}^3 \frac{m_\ell^2}{m_{B_c}^2} \left(1-\frac{m_\ell^2}{m_{B_c}^2}\right)^2,
\end{eqnarray}
in which the  suppression factor $ {m_\ell^2}/{m_{B_c}^2}$ arises from the helicity flip. As a result, the $B_c\to \mu\bar\nu_\mu$ and $B_c\to e\bar\nu_e$ have tiny branching fractions that may be out of the detector  capability at the current experimental facilities.   Secondly, there is only one physical observable, namely the decay rate, and thus the $B_c\to \ell\bar\nu$ is not capable  to uniquely determine all, typically more than one when relativistic corrections are taken into account,  long-distance matrix elements (LDMEs).

On the contrary,  the $B_c\to \gamma \ell\bar\nu$
can provide a wealth of information~\cite{Chang:1997re,Chiladze:1998ny,Lih:1999it,Colangelo:1999gb,Chang:1999gn},  in terms of a number of observables ranging
from the decay probabilities,  polarizations to an angular analysis. It is interesting to notice that the counterpart in $B$ sector,  $B\to \gamma\ell\bar\nu$,  has been widely discussed towards the understanding of the $B$ meson light-cone distribution amplitudes~\cite{Charng:2005fj,Cirigliano:2005ms,Beneke:2011nf,Braun:2012kp,Aubert:2009ya}.
The small branching
fraction  of  $B_c\to \gamma \ell\bar\nu$
can be compensated by the high luminosity at the ongoing hadron colliders and the under-design experimental facilities.
The main purpose of this paper is to explore  the $B_c\to \gamma \ell\bar\nu$  at next-to-leading order (NLO) in $\alpha_s$ and in $|\bold v|^2$, which shall catch up the progress in the $B_c\to \ell\bar\nu$~\cite{Chiladze:1998ny,Lee:2010ts}. For the leptonic decay constant, the two-loop calculation is also available in Ref.~\cite{Onishchenko:2003ui}.

The rest of this paper is organized as follows. In Sec.~\ref{sec:Bctogammalnu_partial_width}, we will derive the formulas for  various partial decay widths of $ B_c\to \gamma \ell\bar\nu$. Sec.~\ref{sec:NLO} is extensively  devoted to the next-to-leading order calculation.   We will discuss the phenomenological results in Sec.~\ref{sec:phenomenolny}.  We summarize our findings and conclude in Sec.~\ref{sec:conclusions}.  We   relegate   the calculation details to the Appendix.

\label{sec:Bctogammalnu_partial_width}

\begin{figure}[t]
\centering
\includegraphics[width=0.9\linewidth]{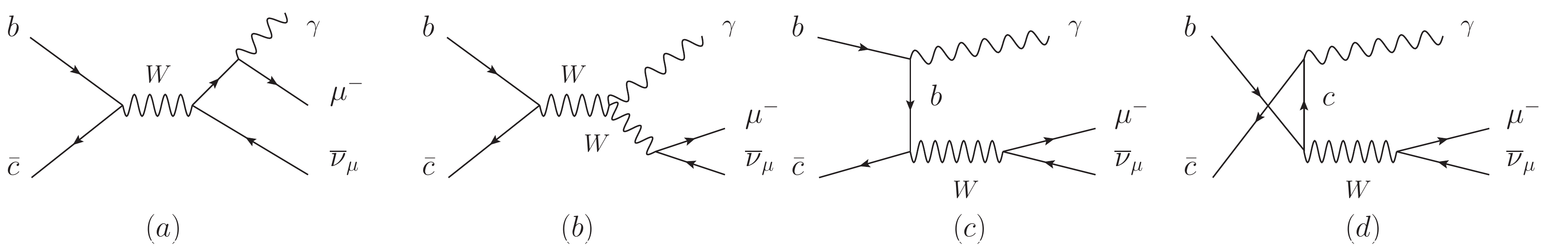}
\caption{Leading order Feynman diagrams for the radiative leptonic
$B_c\to \gamma\mu\bar\nu_\mu$ decay in the SM. The lepton $\mu$ can also be $e$ or $\tau$.  The photon emission from a virtual $W$-boson shown in the second panel   is suppressed by $1/m_W^2$ compared to the other contributions.  }
\label{fig:feynLO}
\end{figure}

In the SM,   leading order (LO) Feynman diagrams for the $B_c\to \gamma\ell\bar\nu$ decay  are shown  in Fig~\ref{fig:feynLO}.   The photon emission from a virtual $W$-boson is suppressed by $1/m_W^2$ compared to   other contributions, and thus the second diagram in Fig.~\ref{fig:feynLO} can be neglected. Integrating out the off-shell $W$-boson, we arrive at the effective  electro-weak Hamiltonian
\begin{eqnarray}
H_{\rm eff} = \frac{G_F}{\sqrt 2} V_{cb} \bar c \gamma_\mu(1-\gamma_5) b \bar l \gamma^\mu(1-\gamma_5)\nu +h.c.,
\end{eqnarray}
where $V_{cb}$ is the CKM matrix element.
The decay amplitude, matrix element of the abo
\section{$ B_c\to \gamma \ell\bar\nu$  }ve Hamiltonian between the $B_c$ and $\gamma\ell\bar\nu$ state,
\begin{eqnarray}
 {\cal A}&=& \langle \gamma l^- \bar\nu| H_{\rm eff}|\overline B_c\rangle
\end{eqnarray}
is responsible for the  process $B_c\to \gamma\ell\bar\nu$.

\subsection{Differential decay widths}

Since there is no  strong  interaction connection  between the leptonic and  hadronic part, the
decay amplitude can be decomposed into  two individual sectors:
\begin{eqnarray}
{\cal A} &=& \frac{G_F}{\sqrt 2} V_{cb}   \bigg\{\langle 0 |\bar c\gamma_\mu(1-\gamma_5) b |\overline B_c\rangle \times  \langle \gamma l^-\bar\nu| \bar l \gamma^\mu(1-\gamma_5)\nu|0\rangle   \nonumber\\
&& \;\;\;\;\;\; \;\;\; + \langle \gamma |\bar c\gamma_\mu(1-\gamma_5) b |\overline B_c\rangle \times  \bar u_l \gamma^\mu(1-\gamma_5)v_{\nu}  \bigg\}, \label{eq:decayAmp}
\end{eqnarray}
with  the  matrix elements encoding the hadronic effects:
\begin{eqnarray}
\langle 0 |\bar c\gamma_\mu(1-\gamma_5) b |\overline B_c\rangle,\;\;\;
\langle \gamma|\bar c\gamma_\mu(1-\gamma_5) b |\overline B_c\rangle.
\end{eqnarray}
The first one defines  the $B_c$ decay constant
\begin{eqnarray}
 \langle 0| \bar c\gamma_\mu\gamma_5 b|\overline B_c (p_{B_c})\rangle &=& i f_{B_c} p_{B_c,\mu},\label{eq:decay_constant_def}
\end{eqnarray}
while the $B_c\to \gamma$ transition is parametrized by two form factors:
\begin{eqnarray}
 \langle \gamma(\epsilon, k)|\bar c\gamma_\mu b |\overline B_c (p_{B_c})\rangle &=& - e\frac{V(L^2)}{p_{B_c}\cdot k}  \epsilon_{\mu\nu\rho\sigma} \epsilon^{*\nu} p_{B_c}^\rho k^\sigma, \label{eq:vector-ff}\\
 \langle \gamma(\epsilon, k)|\bar c\gamma_\mu\gamma_5 b |\overline B_c (p_{B_c})\rangle &=& i eA(L^2) \left(\epsilon_{\mu}^*  -k_\mu \frac{p_{B_c}\cdot \epsilon^*}{p_{B_c}\cdot k}  \right)-\frac{i e}{p_{B_c}\cdot k}f_{B_c} p_{B_c\mu} p_{B_c}\cdot \epsilon^* , \label{eq:axial-vector-ff}
\end{eqnarray}
with the momentum transfer $L= p_{B_c}- k$.  Here and throughout this work we adopt
the convention $\epsilon^{0123}=+1$. The above equations are similar with  the parameterization of  the $B\to \gamma$ form factors as given  in Ref.~\cite{Eilam:1995zv}.   The    last term  in Eq.~\eqref{eq:axial-vector-ff} that is proportional to the $B_c$ decay constant  has been added  in order to  maintain the gauge invariance of the full amplitude~\cite{Grinstein:2000pc,Kruger:2002gf},
and see appendix~\ref{sec:ward_Iden} for a derivation.

Substituting Eqs.~\eqref{eq:decay_constant_def}, \eqref{eq:vector-ff}, \eqref{eq:axial-vector-ff} into Eq.~\eqref{eq:decayAmp}, we obtain
\begin{eqnarray}
{\cal A}
&=& -i\frac{G_F}{\sqrt 2}  V_{cb} e f_{B_c}   \bar u_\ell \gamma^\mu(1-\gamma_5)v_{\nu} \bigg\{[1+a(s_l)]  \left(\epsilon_{\mu}^*  -k_\mu \frac{p_{B_c}\cdot \epsilon^*}{p_{B_c}\cdot k}  \right)- \frac{iv(s_l)}{p_{B_c}\cdot k}    \epsilon_{\mu\nu\rho\sigma} \epsilon^{*\nu} p_{B_c}^\rho k^\sigma     \bigg\},\nonumber\\
\end{eqnarray}
where $s_l=L^2$ and terms due to lepton mass corrections have been neglected. Apparently, this expression is gauge invariant. For the sake of simplicity, we have defined two abbreviations in the above~\footnote{One shall distinguish the form factor $v$ from the relative velocity $\bold{v}$ to be defined in the following.}
\begin{eqnarray}
 a(L^2) \equiv \frac{A(s_l)}{ f_{B_c}}, \;\;\; v(s_l) \equiv \frac{V(s_l)}{f_{B_c}}.
\end{eqnarray}


In terms of the decay constant and form factors, the differential decay width for the $B_c\to \gamma \ell^-\bar\nu_\ell$ is given as
\begin{eqnarray}
 \frac{d^2\Gamma}{dE_k dE_l}  &=&  \frac{1}{ 64 m_{B_c}\pi^3}   |{\cal A}|^2 \nonumber\\
 &=& \frac{\alpha_{\rm em}f_{B_c}^2 | V_{cb} |^2G_F^2m_{B_c}}{4\pi^2x_k^2}(1-x_k)
  \times
   \bigg[a^2 \left(x_k^2+2 x_k (x_l-1)+2 (x_l-1)^2\right)\nonumber\\
 &&+2 a \left((
   v+1) x_k^2+2 (v+1) x_k (x_l-1)+2 (x_l-1)^2\right)+2 v x_k (x_k+2 x_l-2) \nonumber\\
  &&+v^2
   \left(x_k^2+2 x_k (x_l-1)+2 (x_l-1)^2\right)+x_k^2+2 x_k x_l-2 x_k+2
   x_l^2-4 x_l+2\bigg], \nonumber\\
\end{eqnarray}
where $x_k=2E_k/m_{B_c}$ and $y=2E_l/m_{B_c}$, and $E_k$ and $E_l$ is the energy of the photon and charged lepton  in the $B_c$ rest frame, respectively.
One can integrate out the $E_l$ and obtain
\begin{eqnarray}
 \frac{d\Gamma}{dE_k }  =\frac{\alpha_{\rm em}f_{B_c}^2 | V_{cb} |^2G_F^2m_{B_c}^2x_k(1-x_k)((1+a)^2+v^2)}{12\pi^2}.\label{eq:dgammaEk}
\end{eqnarray}

The differential  distributions can also be converted to
\begin{eqnarray}
 \frac{d^2\Gamma}{ds_l d\cos\theta_l}
 &=&  \frac{m_{B_c}^2-s_l}{ 32 m_{B_c}  \pi^2}    | V_{cb} |^2 \alpha_{\rm em} f_{B_c}^2 G_F^2   (1-x_k)
 \frac{1}{x_k^2}
   \bigg[a^2 \left(x_k^2+2 x_k (x_l-1)+2 (x_l-1)^2\right)\nonumber\\
 &&+2 a \left((
   v+1) x_k^2+2 (v+1) x_k (x_l-1)+2 (x_l-1)^2\right)+2 v x_k (x_k+2 x_l-2) \nonumber\\
  &&+v^2
   \left(x_k^2+2 x_k (x_l-1)+2 (x_l-1)^2\right)+x_k^2+2 x_k x_l-2 x_k+2
   x_l^2-4 x_l+2\bigg], \nonumber\\
\end{eqnarray}
using the relation:
\begin{eqnarray}
 E_k &=& \frac{m_{B_c}^2 -s_l}{2m_{B_c}},\\
 E_l &=& \frac{1}{4m_{B_c}}  \left[(m_{B_c}^2+s_l)- (m_{B_c}^2-s_l)\cos\theta_l\right].
\end{eqnarray}
The  $\theta_l$ is the polar angle between the  lepton $\ell$ flight direction   and the opposite direction of the $B_c$ meson in the rest frame of the $\ell\bar\nu_{\ell}$ pair.
Likewise one can integrate out the $\theta_l$
\begin{eqnarray}
 \frac{d\Gamma}{ds_l }  =\frac{\alpha_{\rm em}f_{B_c}^2 | V_{cb} |^2G_F^2(m_{B_c}^2-s_l)s_l((1+a)^2+v^2)}{24\pi^2m_{B_c}^3}.
\end{eqnarray}

\subsection{NRQCD factorization}

The factorization properties for the $B_c\to \gamma \ell\bar\nu$ depend  on the kinematics of the photon.  In this work,  we will not study  the soft-photon contribution as discussed in $B$ decays~\cite{Becirevic:2009aq}, and leave it for future work.  In the region where the photon is a collinear (fast-moving) object, its interaction with   heavy quarks is highly virtual and thus should be encoded in the short distance coefficients.
In the NRQCD scheme, we only need retain those color-singlet operator matrix elements that connect the $B_c$ state to the  vacuum. To the desired order, one expects the following factorization formula:
\begin{eqnarray}
 f_{B_c} &=& \sqrt{\frac{2}{m_{B_c}}}\left[ c_{0}^f \langle 0|\chi^\dagger_c \psi_b |\overline B_c(\textbf{p})\rangle + \frac{c_{2}^f}{m_{B_c}^2} \langle 0|\chi^\dagger_c \left(-\frac{i}{2}  \overleftrightarrow {\bold D}\right)^2  \psi_b |\overline B_c(\textbf{p})\rangle +{\cal O}({\bold v}^4)\right],\label{eq:fBcExpansionInNRQCD}\\
 V &=& \sqrt{\frac{2}{m_{B_c}}}\left[ \frac{c_{0}^V}{m_{B_c}} \langle 0|\chi^\dagger_c \psi_b |\overline B_c(\textbf{p})\rangle + \frac{c_{2}^V}{m_{B_c}^3} \langle 0|\chi^\dagger_c \left(-\frac{i}{2} \overleftrightarrow {\bold D}\right)^2  \psi_b |\overline B_c(\textbf{p})\rangle+{\cal O}({\bold v}^4)\right],~~~~\label{eq:VExpansionInNRQCD}\\
 A &=& \sqrt{\frac{2}{m_{B_c}}}\left[ \frac{ c_{0}^A}{m_{B_c}} \langle 0|\chi^\dagger_c \psi_b |\overline B_c(\textbf{p})\rangle + \frac{c_{2}^A}{m_{B_c}^3} \langle 0|\chi^\dagger_c \left(-\frac{i}{2}  \overleftrightarrow {\bold D}\right)^2  \psi_b |\overline B_c(\textbf{p})\rangle+{\cal O}({\bold v}^4)\right],\label{eq:AExpansionInNRQCD}
\end{eqnarray}
where ${\bold v}$ denotes half relative velocity between the charm and bottom quarks in the meson,  $c_0^{f,V,A}$ and $c_2^{f,V,A}$ are the dimensionless short-distance coefficients that can be expanded  in terms of the strong coupling constant~\footnote{Throughout this paper, we shall use the superscripts $(0)$ and $(1)$ to indicate the LO and NLO contributions in $\alpha_s$ and the subscripts $0$ and $2$ to denote the LO and NLO contributions in the velocity.}.  We shall calculate the one-loop corrections to the  $c_{0}^ {f,V,A}$, but  give only the LO results for $c_2^{f,V,A}$ since the latter ones  are already  power-suppressed.  $\psi_Q$ and $\chi^\dagger_Q$ represent Pauli spinor fields that annihilate the heavy quark $Q$ and anti-quark $\bar{Q}$, respectively. Besides, one need note that the state
 $|H(p)\rangle$ in QCD has the standard normalization:  $\langle H(p^\prime)|H(p)\rangle=2E_p(2\pi)^3\delta^3 (\textbf{p}-\textbf{p}^\prime)$, while an additional factor $2E_p$ is abandoned in the nonrelativistic normalization  where $\langle H(\textbf{p}^\prime)|H(\textbf{p})\rangle=(2\pi)^3\delta^3 (\textbf{p}-\textbf{p}^\prime)$.

\section{Next-to-leading order calculation}
\label{sec:NLO}

\subsection{Kinematics}
Let $p_1$ and $p_2$ represent the momenta for the heavy quark $Q$ and anti-quark $\bar{Q^\prime}$. Without loss of generality, one may adopt the decomposition:
\begin{eqnarray}
p_1 &=&  \alpha \,P_{B_c}-q,\\\
p_2 &=&  \beta\, P_{B_c}+q,
\end{eqnarray}
where $P_{B_c}$ is the total momentum of the quark pair. $q$ is a half of the relative momentum between the
quark pair with $P_{B_c}\cdot q=0$. $\alpha$ and $\beta$ are the energy fraction for  $Q$ and $\bar{Q^\prime}$ in the meson, respectively. The explicit expressions for all the momentum in the rest frame of the $B_c$ meson are given by
\begin{eqnarray}
P_{B_c}^\mu &=&  (E_1+E_2,0),\\
q^\mu &=&  (0,\bold{q}\,),\\
p_1^\mu &=&  (E_1,-\bold{q}\,),\\
p_2^\mu &=&  (E_2,\bold{q}\,).
\end{eqnarray}
In the rest frame, the meson momentum becomes purely timelike while the relative momentum is   spacelike. One  can  obtain the relations $\alpha={\sqrt{m_b^2-q^2}}/({\sqrt{m_b^2-q^2}+\sqrt{m_c^2-q^2}})$  and $\beta=1-\alpha$ with the on-shell conditions $E_1=\sqrt{m_b^2-q^2}$, $E_2=\sqrt{m_c^2-q^2}$, and $q^2=-\bold{q}^2$.

\subsection{Convariant projection method}

In the following calculation, we will adopt  the covariant spin-projector method, which can be applied to all orders in $\bold{v}$.

The Dirac spinors for the $B_c$ system  may be written as
\begin{eqnarray}
u_b(p_1,\lambda) &=&  \sqrt{\frac{E_1+m_b}{2E_1}}\left(
                                           \begin{array}{ll}
                             ~~~~\xi_\lambda \\
\frac{\vec{\sigma}\cdot \overrightarrow{p_1}}{E_1+m_b}\xi_\lambda
                                           \end{array}
                                         \right)\,,
\end{eqnarray}
\begin{eqnarray}
v_c(p_2,\lambda) &=&  \sqrt{\frac{E_2+m_c}{2E_2}}\left(
                                           \begin{array}{ll}
\frac{\vec{\sigma}\cdot \overrightarrow{p_2}}{E_2+m_c}\xi_\lambda\\
                             ~~~~\xi_\lambda\end{array}
                                         \right)\,,
\end{eqnarray}
where $\xi_\lambda$ is the two-component Pauli spinors and $\lambda$ is the polarization parameters. It is straightforward to derive  the covariant form of the spin-singlet combinations
of spinor bilinears:
\begin{eqnarray}
\Pi_0(q) &=&  -i\sum_{\lambda_1,\lambda_2} u_b(p_1,\lambda_1)\bar{v}_c(p_2,\lambda_2)\langle\frac{1}{2}\lambda_1\frac{1}{2}\lambda_2|00\rangle\otimes \frac{\bold{1}_c}{\sqrt{N_c}}\nonumber\\
&=&\frac{i}{4\sqrt{2 E_1 E_2}\omega}(\alpha \,p\!\!\!\slash_{B_c}-q\!\!\!\slash+m_b)\frac{p\!\!\!\slash_{B_c}+E_1+E_2}{E_1+E_2}\gamma_5
(\beta\,p\!\!\!\slash_{B_c}+q\!\!\!\slash-m_c)\otimes \frac{\bold{1}_c}{\sqrt{N_c}}\,,\nonumber\\
\end{eqnarray}
with the auxiliary parameter $\omega=\sqrt{E_1+m_b}\sqrt{E_2+m_c}$.
Here $\bold{1}_c$ is the unit matrix in the fundamental representation of the color SU(3) group.

\subsection{Perturbative matching}

Due to the simplicity of the final state, one can directly match the  QCD currents  onto the NRQCD ones.
To determine the values of $c_0$ and $c_2$, we follow the spirit  that those short-distance coefficients  are insensitive to the long-distance hadronic dynamics. As a convenient choice, one can replace the physical $B_c^-$ meson by
a free $\bar cb $ pair of the quantum number ${}^1S_0^{[1]}$,
so that both the full amplitude, ${\mathcal
A}[\bar cb ({}^1S_0^{[1]})\to\gamma\ell\bar\nu]$, and the NRQCD
operator matrix elements can be directly  accessed in perturbation theory. The
short-distance coefficients $c_i$ can then be solved by equating the
QCD amplitude ${\mathcal A}$ and the corresponding NRQCD amplitude, order by order in $\alpha_s$.   For this purpose, we   introduce a decay constant and two form factors at the free quark level:
\begin{eqnarray}
 \langle 0| \bar c\gamma_\mu\gamma_5 b|\bar cb ({}^1S_0^{[1]})\rangle &=& i \mathbb{f}  g_{\mu 0}, \label{eq:decay_f_NRQCD}\\
 \langle \gamma(\epsilon, k)|\bar c\gamma_\mu b |\bar cb ({}^1S_0^{[1]})\rangle &=& - e\frac{1}{k\cdot p_{B_c}} \mathbb{V}   \epsilon_{\mu\nu\rho\sigma} \epsilon^{*\nu}p_{B_c}^\rho k^\sigma, \label{eq:decay_V_NRQCD} \\
 \langle \gamma(\epsilon, k)|\bar c\gamma_\mu\gamma_5 b |\bar cb ({}^1S_0^{[1]})\rangle &=& i e\mathbb{A}\left(\epsilon_{\mu}^*  -k_\mu \frac{p_{B_c}\cdot \epsilon^*}{p_{B_c}\cdot k}  \right)-i e\frac{1}{p_{B_c}\cdot k}\mathbb {f} p_{B_c\mu} p_{B_c}\cdot \epsilon^*. \label{eq:decay_A_NRQCD}
\end{eqnarray}
Analogous to (\ref{eq:fBcExpansionInNRQCD},\ref{eq:VExpansionInNRQCD},\ref{eq:AExpansionInNRQCD}), one can write down the  matching formula:
\begin{eqnarray}
\mathbb{f}  &=& c_{0}^f \langle 0|\chi^\dagger_c \psi_b |\bar cb ({}^1S_0^{[1]})\rangle + \frac{c_{2}^f}{(m_b+m_c)^2} \langle 0|\chi^\dagger_c \left(-\frac{i}{2}  \overleftrightarrow {\bold D}\right)^2  \psi_b |\bar cb ({}^1S_0^{[1]})\rangle,\\
\mathbb{V}  &=&  \frac{1}{ {m_b+m_c}}\left[c_{0}^V \langle 0|\chi^\dagger_c \psi_b |\bar cb ({}^1S_0^{[1]})\rangle + \frac{c_{2}^V}{(m_b+m_c)^2} \langle 0|\chi^\dagger_c \left(-\frac{i}{2} \overleftrightarrow  {\bold D}\right)^2  \psi_b |\bar cb ({}^1S_0^{[1]})\rangle \right],~~~~~~ \\
\mathbb{A}  &=&  \frac{1}{{m_b+m_c}}\left[c_{0}^A \langle 0|\chi^\dagger_c \psi_b |\bar cb ({}^1S_0^{[1]})\rangle + \frac{c_{2}^A}{(m_b+m_c)^2} \langle 0|\chi^\dagger_c \left(-\frac{i}{2}  \overleftrightarrow  {\bold D}\right)^2  \psi_b |\bar cb ({}^1S_0^{[1]})\rangle \right],
\end{eqnarray}
where  we have adopted the nonrelativistic normalization.

One can organize the full amplitudes defined in Eqs.~(\ref{eq:decay_f_NRQCD},\ref{eq:decay_V_NRQCD},\ref{eq:decay_A_NRQCD})  in powers of the relative momentum between $\bar c$ and $b$,
denoted by ${\bf q}$. To the desired accuracy, one can truncate the series at ${\cal O}({\bf q}^2)$,
with the first two Taylor coefficients. We will  compute both  amplitudes  at LO in $\alpha_s$ in subsection~\ref{sec:tree_level}, and the calculation at NLO in $\alpha_s$ will be conducted in subsection~\ref{subsec:full:QCD:NLO}.

The NRQCD matrix elements encountered in the above equations  are
particularly simple  at LO in $\alpha_s$:
\begin{eqnarray}
& & \langle0\vert\chi^\dagger\psi\vert
\bar cb({}^1S_0^{[1]})\rangle^{(0)} = \sqrt{2N_c},\nonumber\\
& &
\langle0\vert\chi^\dagger(-\frac{i}{2}{\overleftrightarrow{ {\bold D}}})^2\psi\vert
\bar cb({}^1S_0^{[1]}) \rangle^{(0)} = \sqrt{2 N_c}\,{\bf
q}^2,
\label{NRQCD:matrix:elements:LO}
\end{eqnarray}
where the factor $\sqrt{2N_c}$ is due to the spin and color factors
of the normalized $\bar cb({}^1S_0^{[1]})$ state. The
computation of these matrix elements to ${\cal O}(\alpha_s)$ will be
addressed in subsection~\ref{sec:NRQCD:NLO}.

\subsection{Tree-level amplitude}
\label{sec:tree_level}

Adopting the above notation, one can easily obtain the tree-level amplitude for the decay constant
\begin{eqnarray}
 \langle 0| \bar c\gamma_\mu\gamma_5 b|\bar cb ({}^1S_0^{[1]})\rangle^{(0)} &=& {\rm Tr}\left[\Pi_0(q)\gamma_\mu\gamma_5\right]\nonumber\\
 &=& i p_{B_c}^\mu \sqrt{2N_c }\frac{ (E_1+m_b)(E_2+m_c) + q^2}{ 2\sqrt{ E_1E_2(E_1+m_b)(E_2+m_c)} (E_1+E_2)} \nonumber\\
 &=& i g_{\mu0} \sqrt{2N_c} \left(1- \frac{\bold{q}^2}{8m_{red}^2}\right),\label{eq:asymptotic_QCD_f}
 \end{eqnarray}
where the $q^\mu$ terms have been omitted and
\begin{eqnarray}
m_{red}=\frac{m_bm_c}{m_b+m_c}\,,
\end{eqnarray}
is defined as the reduced mass of the $\bar cb$ system.

The vector current is similarly  evaluated as:
\begin{eqnarray}
 \langle \gamma|\bar c\gamma_\mu b |\bar cb ({}^1S_0^{[1]})\rangle^{(0)} &=& {\rm Tr}[ \Pi_0(q)  i ee_c \epsilon\!\!\!\slash^* \frac{i(k\!\!\!\slash-p\!\!\!\slash_2+m_c)}{(k-p_2)^2-m_c^2}\gamma_\mu ]  + {\rm Tr}[ \Pi_0(q) \gamma_\mu\frac{i(p\!\!\!\slash_1-k\!\!\!\slash+m_b)}{(p_1-k)^2-m_b^2}  i ee_b \epsilon\!\!\!\slash^* ] \nonumber\\
 &=&     -\frac{ e   \sqrt{2N_c}  }{4w\sqrt{E_1E_2 }  } (\frac{  e_c  }{ E_2k\cdot p_{B_c} + Ek\cdot q}+ \frac{  e_b  }{ E_1k\cdot p_{B_c} - Ek\cdot q})
 \nonumber\\
 && \times \bigg\{  E_{bc}\epsilon_{\mu\nu\rho\sigma} \epsilon^{*\nu} k^\rho p_{B_c}^\sigma + E(E_1+E_2+m_b-m_c) \epsilon_{\mu\nu\rho\sigma} \epsilon^{*\nu} k^\rho q^\sigma \bigg\}\label{eq:asymptotic_QCD_V}.
\end{eqnarray}
We have  introduced the abbreviation $E=E_1+E_2$, and $E_{bc}=(E_1+m_b)(E_2+m_c) +q^2$.
Here $e_c= 2/3$ and $e_b=-1/3$ is the electric charge of the $c$ and $b$ quark, respectively.

One can perform the  Taylor expansion of the amplitudes in powers of $q^\mu$:
\begin{eqnarray}
{\cal A}(q)&=& {\cal A}(0)+\frac{\partial {\cal A}(0)}{\partial q^\mu}\mid_{q=0}q^\mu+\frac{1}{2!}\frac{\partial^2 {\cal A}(0)}{\partial q^\mu\partial q^\nu}\mid_{q=0}q^\mu q^\nu+\ldots.
\end{eqnarray}
Those terms linear in $q$ should be dropped since this auxiliary momentum introduced at the quark level has no correspondence at the hadron level.
In this paper,  the ${\cal O}(|\bold{q}|^2)$ contributions  will be retained.   In order to simplify the calculation  in  the covariant derivation, one shall use the following replacement:
\begin{eqnarray}
q^\mu q^\nu&\rightarrow& \frac{|\bold{q}|^2}{D-1}(-g^{\mu\nu}+\frac{P_{B_c}^\mu P_{B_c}^\nu}{P_{B_c}^2}).
\end{eqnarray}

The result for the axial-vector current is a bit lengthy:
 \begin{eqnarray}
 \langle \gamma|\bar c\gamma_\mu\gamma_5 b |\bar cb ({}^1S_0^{[1]})\rangle^{(0)}
 &=& -i e \sqrt{2N_c}\frac{ 1} {4\sqrt{E_1E_2(E_1+m_b)(E_2+m_c)}} \nonumber\\
 &&\times \bigg\{ \epsilon_{\mu}^*  e_c\frac{ k\cdot p_{B_c} E_{bc}+k\cdot q E(E_1-E_2+m_b-m_c)} { E_2 k\cdot p_{B_c}+E k\cdot q} \nonumber\\
 &&  - \epsilon_{\mu}^*  e_b\frac{ k\cdot p_{B_c} E_{bc} +k\cdot q E(E_1-E_2+m_b-m_c)} { E_1 k\cdot p_{B_c}-E k\cdot q} \nonumber\\
 && + q_\mu e_c  \frac{2 (E_1-E_2+m_b-m_c)(E_2\epsilon^*\cdot p_{B_c}+E\epsilon^*\cdot q)}{E_2 k\cdot p_{B_c}+Ek\cdot q}\  \nonumber\\
 && - q_\mu e_b  \frac{2 (E_1-E_2+m_b-m_c)(E_1\epsilon^*\cdot p_{B_c}-E\epsilon^*\cdot q)}{E_1 k\cdot p_{B_c}-Ek\cdot q} \nonumber\\
&& + p_{B_c \mu}   e_c\frac{2E_{bc}(E_2\epsilon^*\cdot p_{B_c}  +E\epsilon^*\cdot q)} {E(E_2 k\cdot p_{B_c}+E k\cdot q)} \nonumber\\
&& -p_{B_c \mu}   e_b\frac{2(E_1E_{bc}\epsilon^*\cdot p_{B_c}  +E\epsilon^*\cdot q(E_{bc}+q^2)} {E(E_1 k\cdot p_{B_c}-E k\cdot q)}  \nonumber\\
&& -k_{\mu}   e_c\frac{E_{bc}\epsilon^*\cdot p_{B_c}  +E\epsilon^*\cdot q(E_1-E_2+m_b-m_c)} { E_2 k\cdot p_{B_c}+E k\cdot q} \nonumber\\
&& +k_{\mu}   e_b\frac{E_{bc}\epsilon^*\cdot p_{B_c}  +E\epsilon^*\cdot q(E_1-E_2+m_b-m_c)} { E_1 k\cdot p_{B_c}-E k\cdot q} \bigg\}  \label{eq:asymptotic_QCD_A}.
\end{eqnarray}
In order to extract the ${\cal A}$ form factor, we  only need to keep the $\epsilon_\mu$ term which corresponds to Feynman gauge $\epsilon\cdot p_{B_c}=0$, but we have explicitly checked the gauge invariance up to $\bold{v}^2$ order.


The tree-level NRQCD matrix elements for the $\bar cb$ have  been given in Eq.~(\ref{NRQCD:matrix:elements:LO}), and thus the  above  results in Eqs.~(\ref{eq:asymptotic_QCD_f},\ref{eq:asymptotic_QCD_V},\ref{eq:asymptotic_QCD_A})  lead to  the tree-level Wilson coefficients
\begin{eqnarray}
 c_{0}^{f,0}&=&1, \\
 c_{2}^{f,0} &=& -\frac{\tilde{z}^4}{8z^2},\\
 c_{0}^{V,0}&=&   -\frac{e_c}{2z}-  \frac{e_b}{2}  ,\\
 c_{2}^{V,0}&=&  -\tilde{z}^2 \left( \frac{e_c(3z^2+2z+11)}{48 z^3} + \frac{e_b(11z^2+2z+3)}{48 z^2}\right),\\
 c_{0}^{A,0}&=&   \frac{e_b}{2} - \frac{e_c}{2z},
  \end{eqnarray}
 \begin{eqnarray}
 c_{2}^{A,0}&=&  -\tilde{z}^2 \bigg( \frac{e_c[(3z^2+2z+11)+ 8 z(1-z) m_b/E_k]}{48 z^3} \nonumber\\
 &&  - \frac{e_b[(11z^2+2z+3)-  8 z(1-z) m_b/E_k]}{48 z^2}\bigg).
\end{eqnarray}
In the above results,  we have defined $z= m_c/m_b$ and $\tilde{z}=1+z$.  $c_{i}^{f,0}$ means the LO of Wilson coefficient $c_{i}^f$. It is interesting to notice that the Wilson coefficients $ c_{2}^{A,0}$ depends on the energy of the emitted photon, which will  induce  nontrivial behaviors as  will be demonstrated later.

\subsection{NLO amplitudes in QCD}
\label{subsec:full:QCD:NLO}
\begin{figure}
\centering
\includegraphics[width=0.9\textwidth]{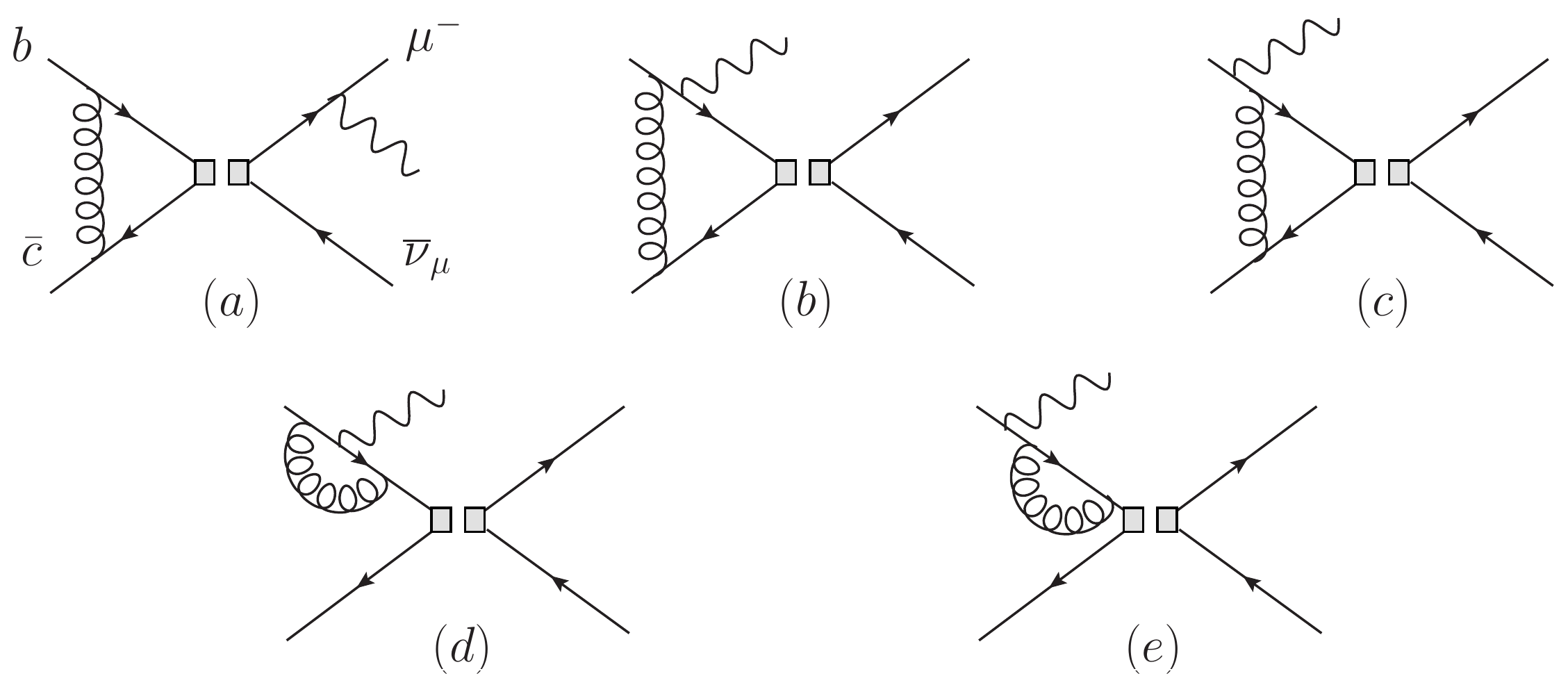}
\caption{ Typical NLO Feynman diagrams for the radiative leptonic  $B_c\to \gamma\mu\bar\nu_\mu$ decay in the SM. The other four diagrams can be easily obtained  by interchanging the bottom and anti-charm quarks lines. }
\label{fig:feynNLO}
\end{figure}

Typical one-loop diagrams for the QCD corrections to the $B_c\to \gamma\ell\bar{\nu}_\ell$ decay are shown  in Fig.~\ref{fig:feynNLO}.
In calculating  the one-loop amplitudes, we use the dimensional regularization to regulate  the ultraviolet (UV) and infrared (IR) divergence.

The   diagram (a) in Fig.~\ref{fig:feynNLO} contributes to the NLO decay constant:\begin{eqnarray}
\mathbb {f}^{1}_{0,a} = \sqrt{2N_c} \frac{C_F \alpha_s}{4\pi}\bigg[\frac{1}{\hat \epsilon_{UV}}+\frac{2}{\hat \epsilon_{IR}}+3\ln\frac{\mu^2}{m_b^2} -2 +2t_1-\frac{6\ln z}{ z+1}\bigg],
\end{eqnarray}
with
\begin{eqnarray}
 t_1 &=& \frac{1}{ {2|\bold{v}|}} \left(\pi^2 - i\pi \left[\frac{1}{\hat \epsilon_{IR}} -\ln \frac{16m_{red}^2|\bold{v}|^2}{\mu^2}\right]\right),\nonumber\\
\bold v & =& \frac{\bold q}{2m_{red}}.
\end{eqnarray}
We  have introduced the abbreviation
\begin{eqnarray}
\frac{1}{\hat \epsilon_{UV,IR}}=\frac{1}{\epsilon_{UV,IR}}-\gamma_E+\ln 4\pi.
\end{eqnarray}

The heavy quark field renormalization  and mass term are given as
\begin{eqnarray}
Z^{OS}_{q} & = &1-\frac{C_F \alpha_s}{4\pi}\left[\frac{1}{\hat\epsilon_{UV}}+\frac{2}{\hat\epsilon_{IR}}+3\ln\frac{\mu^2}{m^2}+4\right],\nonumber\\
\delta_{m}&=& -\frac{3mC_F\alpha_s }{4\pi} \bigg[\frac{1}{\hat\epsilon_{UV}} +\ln \frac{\mu^2}{m^2} +\frac{4}{3}\bigg].
\end{eqnarray}

For the vector current form factor, the sub-diagram in Fig.~\ref{fig:feynNLO} gives out the corresponding contribution
\begin{eqnarray}
\mathbb{V}_{b} &=& \frac{\sqrt{2N_c}e_b C_F \alpha_s}{4\pi m_b}[-\frac{1}{\hat \epsilon_{IR}}+\frac{4 \tilde{z}}{y^2-\tilde{z}^2}+\frac{\tilde{z}^2+y^2}{y^2 \tilde{z}-\tilde{z}^3}b_1-\frac{1}{\tilde{z}}b_2+\frac{2 \left(y^2-z (z+1)\right)}{z \left(y^2-\tilde{z}^2\right)}b_3\nonumber\\&&-\frac{2 y^2}{z \left(y^2-\tilde{z}^2\right) }b_4 -\frac{y^2}{  \tilde{z}}c_4 -(1-z)c_3-(\tilde{z}^2-y^2)d_1],\nonumber\\
\mathbb{V}_{c} &=&  \frac{\sqrt{2N_c}e_b C_F \alpha_s}{4\pi m_b}[-\frac{1}{2}\frac{1}{\hat \epsilon_{UV}}+\frac{y^2+z^2+4 z+3}{\tilde{z}^2-y^2}+\frac{\tilde{z}+y^2}{\tilde{z}^2-y^2}b_1-\frac{\tilde{z} \left(3 y^2-z^2+1\right)}{2 z \left(\tilde{z}^2-y^2\right)}b_4\nonumber\\
&&+\frac{\left(2 z^2+3 z-1\right) \tilde{z}-y^2 (2 z+3)}{2 z \left(y^2-\tilde{z}^2\right)}b_3+({\tilde{z}+y^2-z^2}) c_4],\nonumber\\
\mathbb{V}_{d} &=&  \frac{\sqrt{2N_c}e_b C_F \alpha_s}{4\pi m_b}[-\frac{1}{2}\frac{1}{\hat \epsilon_{UV}}+\frac{y^2-z^2+4 z+5}{\tilde{z}^2-y^2}+\frac{y^2-z^2+z+2}{\tilde{z}^2-y^2}b_2\nonumber\\
&&+\frac{y^2-z^2+4 z+5}{2 (y^2- \tilde{z}^2)}b_3+c_1],\nonumber\\
\mathbb{V}_{e} &=&  \frac{\sqrt{2N_c}e_b C_F \alpha_s}{4\pi m_b}[\frac{-y^2+z^2+8 z+7}{2(\tilde{z}^2-y^2)}+\frac{\tilde{z}^2-y^2}{2 \left(y^2-z \tilde{z}\right)}-\frac{\tilde{z} \left(y^2-z^2+1\right)}{2 \left(y^2-z \tilde{z}\right) \left(y^2-\tilde{z}^2\right)}b_2 \nonumber\\
&& +\frac{\left(z^2+6 z+1\right)
   \tilde{z}^2+y^4-2 y^2 \left(z^2+4 z+3\right)}{2 (y^2- \tilde{z}^2) \left(y^2-z \tilde{z}\right)}b_3],
\end{eqnarray}
where the auxiliary functions $b_i$, $c_i$, and $d_i$ are defined    in Appendix~\ref{ap-Integral}.

The counter-mass terms and wave function renormalization corrections give:
\begin{eqnarray}
\mathbb{V}_{CT-m} &=&\frac{\sqrt{2N_c}e_b C_F \alpha_s}{4\pi m_b}[\frac{3 \tilde{z}}{y^2-\tilde{z}^2}(\frac{1}{\hat \epsilon_{UV}}+\ln\frac{\mu^2}{m_b^2}+\frac{4
   }{3})],
\nonumber\\
\mathbb{V}_{CT-F} &=&\frac{\sqrt{2N_c}e_b C_F \alpha_s}{4\pi m_b}[\frac{1}{\hat \epsilon_{IR}}+\frac{1}{2}\frac{1}{\hat \epsilon_{UV}}+ \frac{3}{2}\ln\frac{\mu^2}{zm_b^2} +2].
\end{eqnarray}

For the axial-vector current form factor, the sub-diagram has gauge-dependent contributions, however,
the  summed result is gauge-invariant. We will show the detail in   Appendix~\ref{ap-A}.

\subsection{NLO amplitudes in NRQCD}
\label{sec:NRQCD:NLO}
The NRQCD Lagrangian can be derived by integrating out the degrees of freedom of order heavy quark mass~\cite{Bodwin:1994jh}:
\begin{eqnarray}
{\mathcal L}_{\rm NRQCD} &=&
\psi^\dagger \left( i D_t + {{\bf D}^2 \over 2m} \right) \psi+ \psi^\dagger {{\bf D}^4 \over 8m^3} \psi
+ {c_F \over 2 m} \psi^\dagger \bfsigma \cdot g_s {\bf B} \psi
\nonumber\\
&+& {c_D\over 8 m^2} \psi^\dagger ({\bf D}\cdot g_s {\bf E}- g_s {\bf E}\cdot {\bf D})\psi
+{i c_S\over 8 m^2} \psi^\dagger \bfsigma \cdot ({\bf D}\times g_s {\bf E}- g_s {\bf E}\times {\bf D})\psi
\nonumber\\
&+& \left(\psi \rightarrow i \sigma ^2 \chi^*, A_\mu \rightarrow - A_\mu^T\right) +
{\mathcal L}_{\rm light} \,.
\label{NRQCD:Lag}
\end{eqnarray}
The replacement in the last line implies that the corresponding
heavy anti-quark bilinear sector can be obtained through the charge conjugation transformation. ${\mathcal L}_{\rm light}$ represents the Lagrangian for the light quarks and gluons.
The coefficients $c_D$, $c_F$, and $c_S$ have perturbative expansions in powers
of $\alpha_s$, which can be written as $c_i=1+{\cal O}(\alpha_s)$.



The matrix element of the $\bar cb$ to vacuum at NLO can be written as
\begin{eqnarray}
\langle0\vert\chi^\dagger\psi\vert
\bar cb({}^1S_0^{[1]})\rangle^{(1)} =  \sqrt{2N_c}\frac{\alpha_sC_F}{2\pi\, {2|\bold{v}|}}  \left(\pi^2 - i\pi \left[\frac{1}{\epsilon_{IR}} -\ln \frac{16m_{red}^2|\bold{v}|^2}{\mu^2}\right]\right).
\end{eqnarray}
This is in agreement with the results in Ref.~\cite{Jia:2011ah}.

\subsection{Determination of $c_i$: Matching QCD to NRQCD}

Up to ${ \alpha}_s$ and $\bold{v}^2$, one can expand the decay constant and  form factors as
\begin{eqnarray}
\mathbb {f}  &=& c_{0}^{f,0} \langle 0|\chi^\dagger_c \psi_b |\bar cb ({}^1S_0^{[1]})\rangle^{(0)} + c_{0}^{f,1} \langle 0|\chi^\dagger_c \psi_b |\bar cb ({}^1S_0^{[1]})\rangle^{(0)} +  c_{0}^{f,0} \langle 0|\chi^\dagger_c \psi_b |\bar cb ({}^1S_0^{[1]})\rangle^{(1)}\nonumber\\&&+ \frac{c_{2}^{f,0}}{(m_b+m_c)^2} \langle 0|\chi^\dagger_c (-\frac{i}{2}  \overleftrightarrow D)^2  \psi_b |\bar cb ({}^1S_0^{[1]})\rangle^{(0)},\\
\mathbb {V}  &=&  \frac{1}{ {m_b+m_c}}[c_{0}^{V,0} \langle 0|\chi^\dagger_c \psi_b |\bar cb ({}^1S_0^{[1]})\rangle^{(0)} +c_{0}^{V,1} \langle 0|\chi^\dagger_c \psi_b |\bar cb ({}^1S_0^{[1]})\rangle^{(0)}\nonumber\\&&+c_{0}^{V,0} \langle 0|\chi^\dagger_c \psi_b |\bar cb ({}^1S_0^{[1]})\rangle^{(1)}+ \frac{c_{2}^{V,0}}{(m_b+m_c)^2} \langle 0|\chi^\dagger_c (-\frac{i}{2} \overleftrightarrow D)^2  \psi_b |\bar cb ({}^1S_0^{[1]})\rangle^{(0)} ],\\
\mathbb {A}  &=&  \frac{1}{{m_b+m_c}}[c_{0}^{A,0} \langle 0|\chi^\dagger_c \psi_b ^{(0)}|\bar cb ({}^1S_0^{[1]})\rangle^{(0)}+c_{0}^{A,1} \langle 0|\chi^\dagger_c \psi_b ^{(0)}|\bar cb ({}^1S_0^{[1]})\rangle^{(0)}\nonumber\\&&+c_{0}^{A,0} \langle 0|\chi^\dagger_c \psi_b ^{(0)}|\bar cb ({}^1S_0^{[1]})\rangle^{(1)} + \frac{c_{2}^{A,0}}{(m_b+m_c)^2} \langle 0|\chi^\dagger_c (-\frac{i}{2}  \overleftrightarrow D)^2  \psi_b |\bar cb ({}^1S_0^{[1]})\rangle^{(0)} ].
\end{eqnarray}
Matching the QCD results onto the NRQCD, one can obtain the UV and IR finite short-distance coefficient
\begin{eqnarray}
c_{0}^{f,1}= -\frac{ 3C_F \alpha _s }{4 \pi} \left(2 + \frac{1-z}{1+z}\ln z\right),\label{c0f1}
\end{eqnarray}

\begin{eqnarray}
c_{0}^{V,1}&=&\frac{ C_F \alpha _s }{4\pi}\{e_b[\ln \frac{\mu^2}{z m_b^2}-\frac{\tilde{z}^2 \left(-3 z \tilde{z}+\tilde{z}+2 y^2\right)+y^4}{2 \left(y^2-z \tilde{z}\right) \left(y^2-\tilde{z}^2\right)}+\frac{\tilde{z}^3+y^2 (3 z-1)}{4 \left(\tilde{z}^3-y^2 \tilde{z}\right)}b_1+\frac{y^2-2 z \tilde{z}}{2 z \left(y^2-z \tilde{z}\right)}b_3\nonumber\\&&+\frac{1}{4} (\frac{2 \tilde{z}}{y^2-z \tilde{z}}+\frac{2}{\tilde{z}-y}+\frac{2}{\tilde{z}+y}-\frac{4}{\tilde{z}}-3)b_2+\frac{-z \tilde{z}^2+\tilde{z}^2+3 y^2 z-y^2}{2 z \left(y^2-\tilde{z}^2\right)}b_4\nonumber\\&&+\frac{-\tilde{z}-y^2 z+z^3+z^2}{z \tilde{z}}c_1+\frac{y^2 z-z^3+2 z+1}{z}c_2+(z-1)c_3+(y^2-\tilde{z}^2)d_1]\nonumber\\&&+(e_b\to \frac{e_c}{z}, z\to\frac{1}{z},y\to\frac{y}{z})\},\label{c0v}
\end{eqnarray}
\begin{eqnarray}
c_{0}^{A,1}&=&\frac{ C_F \alpha _s }{4\pi}\bigg\{e_b\bigg[-\ln \frac{\mu^2}{z m_b^2}+\frac{1}{2 \left(y^2-z \tilde{z}\right) \left(y^2-\tilde{z}^2\right)^2}(y^4 (z+11) \tilde{z}-y^2 (z (5 z+34)+5) \tilde{z}^2\nonumber\\
&&+(z (z (3 z+23)+5)+1) \tilde{z}^3+y^6)+\frac{b_1}{4 \tilde{z} \left(y^2-\tilde{z}^2\right)^2}(-2 y^2 (z-3) \tilde{z}^2-(z^2+14 z\nonumber\\
&&-3) \tilde{z}^3+y^4 (3 z-1))+\frac{b_2}{4 \tilde{z} \left(y^2-z \tilde{z}\right) \left(y^2-\tilde{z}^2\right)}(y^2 (y^2 (3 z+7)-(2 z+3) (3 z-1) \tilde{z})\nonumber\\
&&+(3 (z-1) z-2) \tilde{z}^3)-\frac{b_3}{2 z (z\tilde{z}-y^2) (y^2-\tilde{z}^2)^2}(y^2 \left(13 z^2-2 z+1\right) \tilde{z}^2-2 (3 z^3+z) \tilde{z}^3\nonumber\\
&&+y^4 (y^2-8 z^2-6 z+2))-\frac{(z-1)^2 \tilde{z}^3+y^4 (3 z+1)-2 y^2 \left(2 z^3+5 z^2+2 z-1\right)}{2 z \left(y^2-\tilde{z}^2\right)^2}b_4\nonumber\\
&&+\frac{y^2 \left(y^2 (-z)+z^2 (2 z+5)-3\right)-(z-1) (z (z+4)-1) \tilde{z}^2}{z \tilde{z} \left(\tilde{z}^2-y^2\right)}c_1\nonumber\\
&&-\frac{((z-2) z (z+4)+1) \tilde{z}^2+y^2 \left(z \left(y^2-2 z (z+2)+3\right)+3\right)}{z \left(y-\tilde{z}\right) \left(\tilde{z}+y\right)}c_2\nonumber\\
&&+\frac{(z-1) \left(-y^2+z^2-1\right)}{y^2-\tilde{z}^2}c_3+(-y^2+z^2+4 z-1)d_1\bigg]\nonumber\\
&&-(e_b\to \frac{e_c}{z}, z\to\frac{1}{z},y\to\frac{y}{z})\bigg\}.\label{c0a}
\end{eqnarray}
Note that the scale dependent term in the brace of Eqs.~(\ref{c0v}) and (\ref{c0a}) will be cancelled each other, the residual dependence only lies in the strong coupling constant.

\section{Phenomenolnical Results}
\label{sec:phenomenolny}

\begin{figure}
\centering
\includegraphics[width=0.48\linewidth]{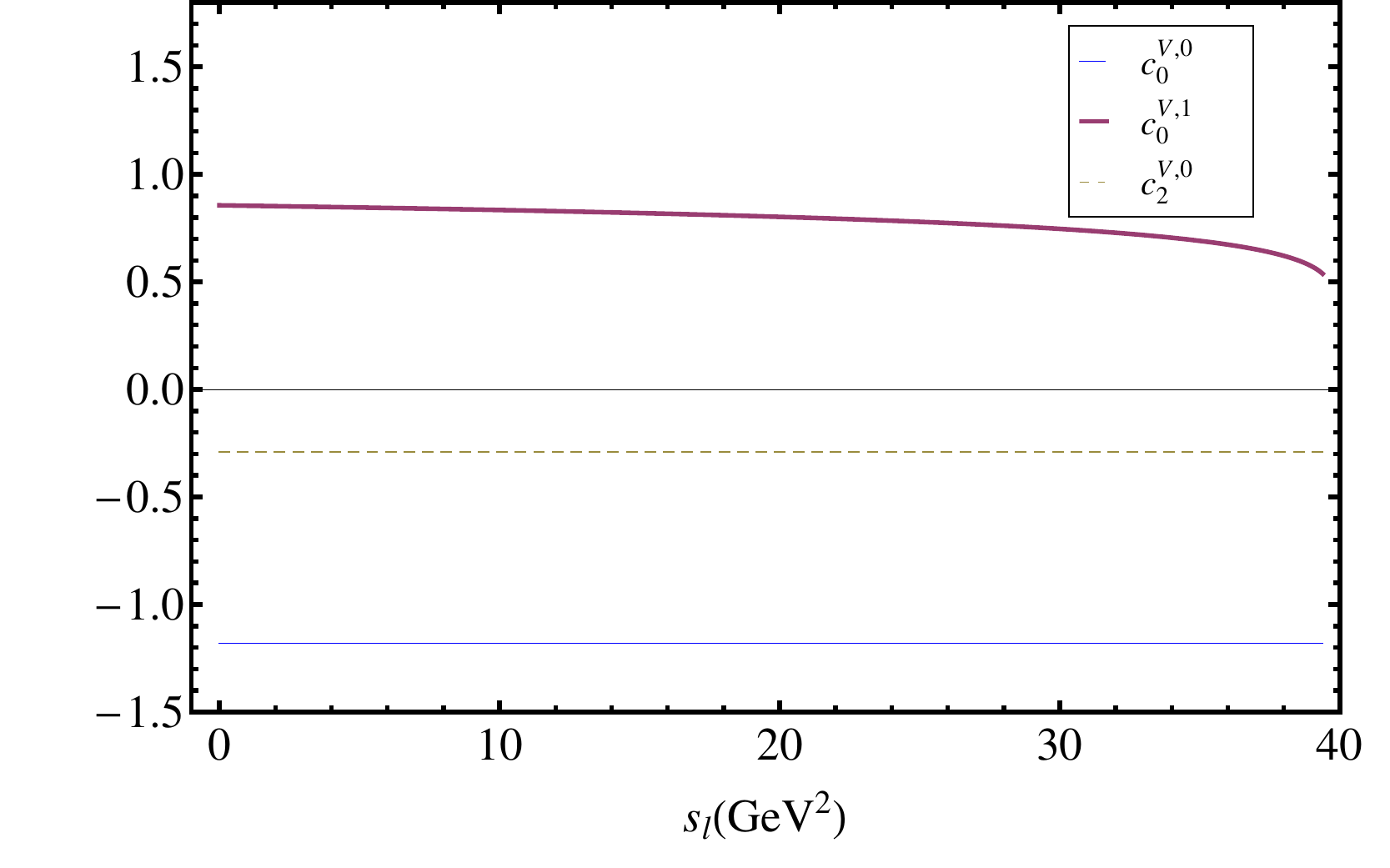}
\includegraphics[width=0.48\linewidth]{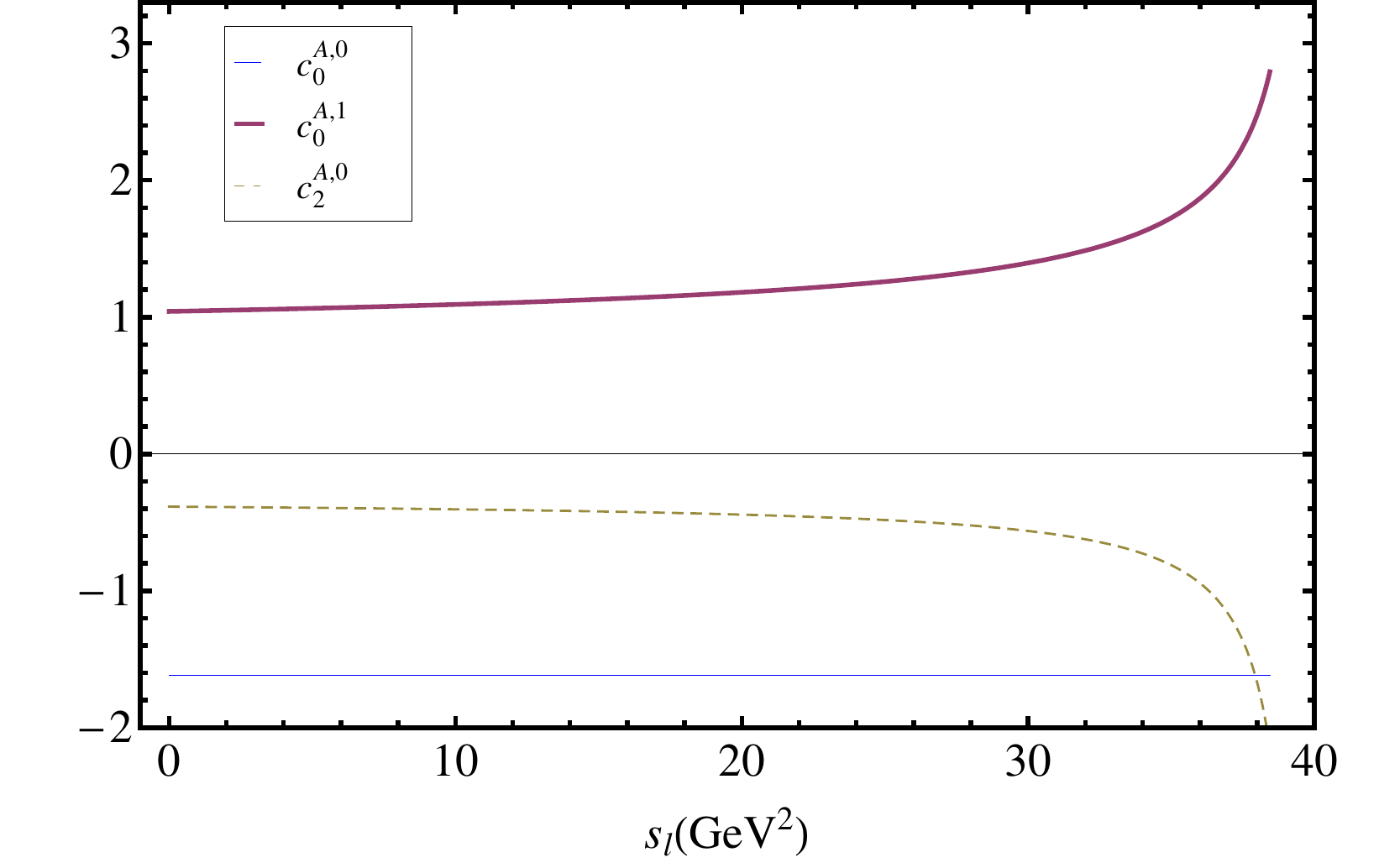}
\caption{Dependence of short-distance coefficients $c^{V(A)}$ on the  $s_l$. The solid line denotes the coefficient $c_0^{V(A),0}$,
the dotted line is the coefficient $c_2^{V(A),0}$ from relativistic corrections, and the thick  curve
is the coefficient $c_0^{V(A),2}$ from $\alpha_s$ corrections.   }
\label{fig:wilson_coeffi_V}
\end{figure}

The input parameters are adopted as~\cite{Agashe:2014kda}:  $m_{B_c}=6.2756$GeV; $G_F=1.16637\times10^{-5}{\rm  GeV^{-2}}$; $\alpha=1/128$; for the CKM parameters, we adopt $|V_{cb}|=0.041$. For the heavy quark mass, we adopt $m_b=4.8$GeV and $m_c=1.5$GeV~\cite{Qiao:2012hp}. The $B_c$-meson lifetime is using the latest measurement by the LHCb Collaboration, i.e. $\tau_{B_c}=0.50{\rm ps}$~\cite{Aaij:2014gka,Aaij:2014bva}.

We first present numerical  results  for  the  decay constant $f_{B_c}$:
\begin{eqnarray}
c_2^{f,0} &=& - \frac{\tilde z^4}{8z^2} = -3.8,\nonumber\\
c_{0}^{f,1} &=& -\frac{ 3C_F \alpha _s }{4 \pi} \left(2 + \frac{1-z}{1+z}\ln z\right) =-0.44\times \alpha_s.
\end{eqnarray}
The strong coupling constant at the Z-boson peak is~\cite{Agashe:2014kda}
\begin{eqnarray}
\alpha_s(m_Z)= 0.1185\pm 0.0006,
\end{eqnarray}
which corresponds to
\begin{eqnarray}
\alpha_s(m_b)= 0.218,\;\; \alpha_s(m_c)=0.368.
\end{eqnarray}
With these values, one can see the $\alpha_s$   corrections can reduce the decay constant by approximately $9.5\%-16.2\%$.

To estimate the size of  ${\cal O}(|\bold v|^2)$ effects, one requests  the size of non-perturbative LDMEs,  for which we use Buchm\"uller-Tye (B-T) potential model~\cite{Buchmuller:1980su}:
\begin{eqnarray}
\langle 0|\chi^\dagger_c \psi_b |\overline B_c(\textbf{p})\rangle&=& \sqrt{\frac{N_{c}}{2\pi}}|R_{S}^{\mbox{\footnotesize B-T}}(0)| \simeq 0.884 {\rm GeV}^{3/2} \,,  \\
\langle 0|\chi^\dagger_c \left(-\frac{i}{2}  \overleftrightarrow {\bold D}\right)^2  \psi_b |\overline B_c(\textbf{p})\rangle&\simeq& \mathbf{q}^{2} \langle 0|\chi^\dagger_c \psi_b |\overline B_c(\textbf{p})\rangle\,.
\end{eqnarray}

For an estimate of $ \mathbf{q}^{2}$, one may make use of the relative velocity. Using the heavy quarks kinetic and potential
energy approximation~\cite{Bodwin:1994jh}, we have
\begin{eqnarray} |\mathbf{  v}| &\simeq&\alpha_s(2m_{red}|\mathbf{  v}| \,)\,.
\end{eqnarray}
Choosing  $m_b=4.8$ GeV and $m_c=1.5$ GeV, and using two-loop strong coupling constant, we get
\begin{eqnarray}
| \mathbf{v}|^2
_{J/\psi}&\approx&0.267\,,~~~~
| \mathbf{v}|^2
_{\Upsilon}\approx0.108\,,~~~
| \mathbf{v}|^2
_{B_c}\approx0.186\,.
\end{eqnarray}

For a
value $\langle \mathbf{v}^2 \rangle_{B_c}\simeq0.186$, we have
\begin{eqnarray}
\mathbf{q}^{2}  \simeq0.9718{\rm GeV}^{2}.
\end{eqnarray}
As a result, the decay constant will be further reduced by about $9\%$.

For the short-distance coefficients for $B_c\to \gamma$ transition form factors $V$ and $A$,
our results are shown in  Fig.~\ref{fig:wilson_coeffi_V}. The solid line denotes the leading-order coefficient $c_0^{V(A),0}$,
the dotted line correspond to the coefficient $c_2^{V(A),0}$ from relativistic corrections, and the thick  curve
is the coefficient $c_0^{V(A),2}$ from $\alpha_s$ corrections.
From these figures, one can see the relativistic corrections give constructive contributions, but
the ${\cal O}( \alpha_s)$ QCD corrections are  destructive and thus have important consequences.
Note that the factorization in Eqs.~(\ref{eq:fBcExpansionInNRQCD},\ref{eq:VExpansionInNRQCD},\ref{eq:AExpansionInNRQCD})
is valid only for a hard photon, while the soft-photon contribution needs special treatment~\cite{Becirevic:2009aq}.  Thus a cut-off on the photon energy should be introduced, however we have checked that
the cut-off will  not affect   the results significantly in Tabs.~\ref{tab:results} and \ref{tab:results2}.

\begin{figure}
\centering
\includegraphics[width=0.48\linewidth]{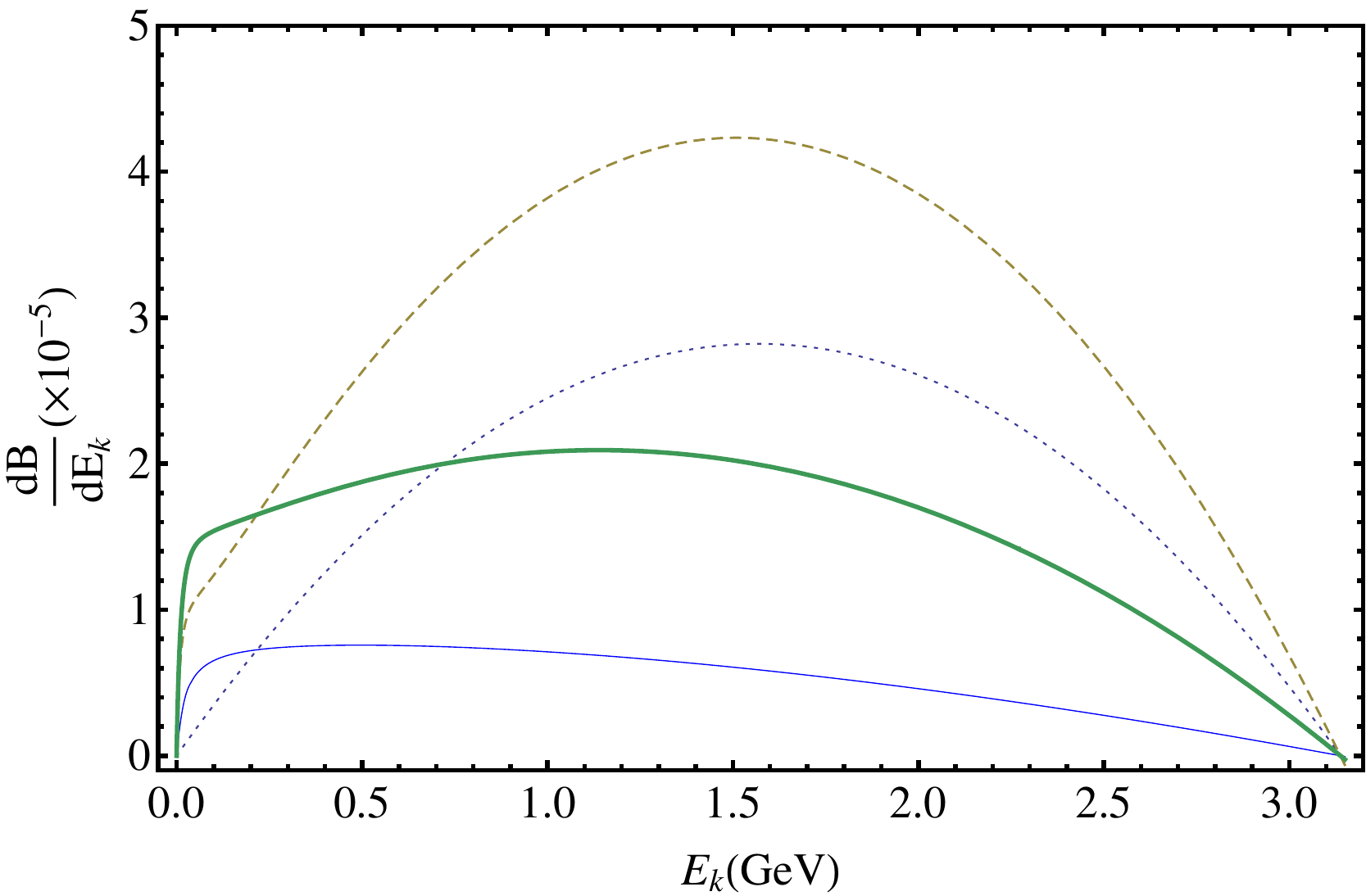}
\includegraphics[width=0.48\linewidth]{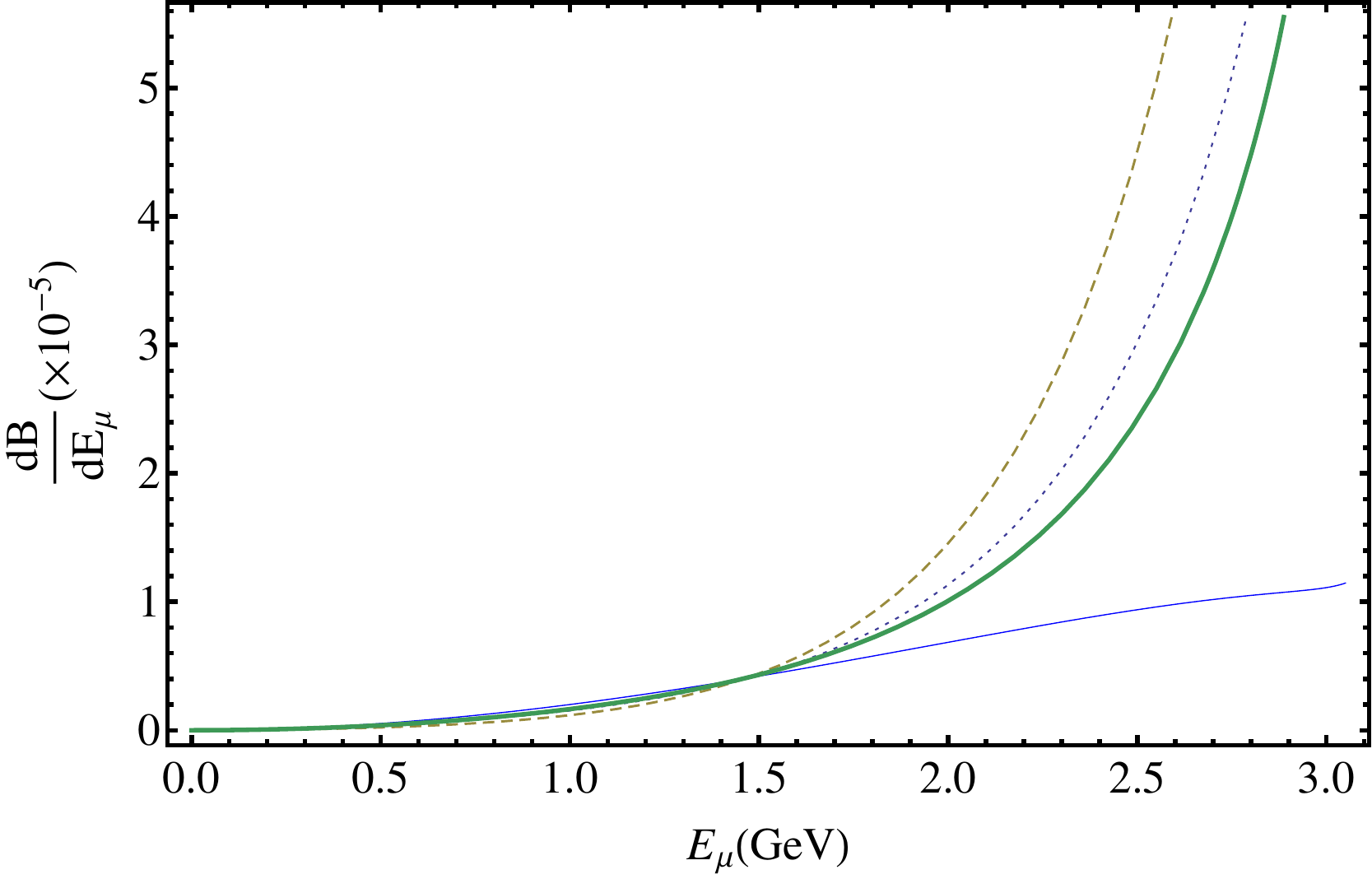}
\caption{The dependence of the branching ratio  ${\cal B}(B_c\to \gamma\mu\bar\nu_{\mu})$ on the photon and lepton energy. The dotted line denotes the leading-order result,
the dashed line is the result with relativistic corrections, the blue line is the result with QCD corrections, and the thick  curve denotes the total results with both the QCD and relativistic  corrections.  }
\label{fig:ek}
\end{figure}
\begin{figure}
\centering
\includegraphics[width=0.7\linewidth]{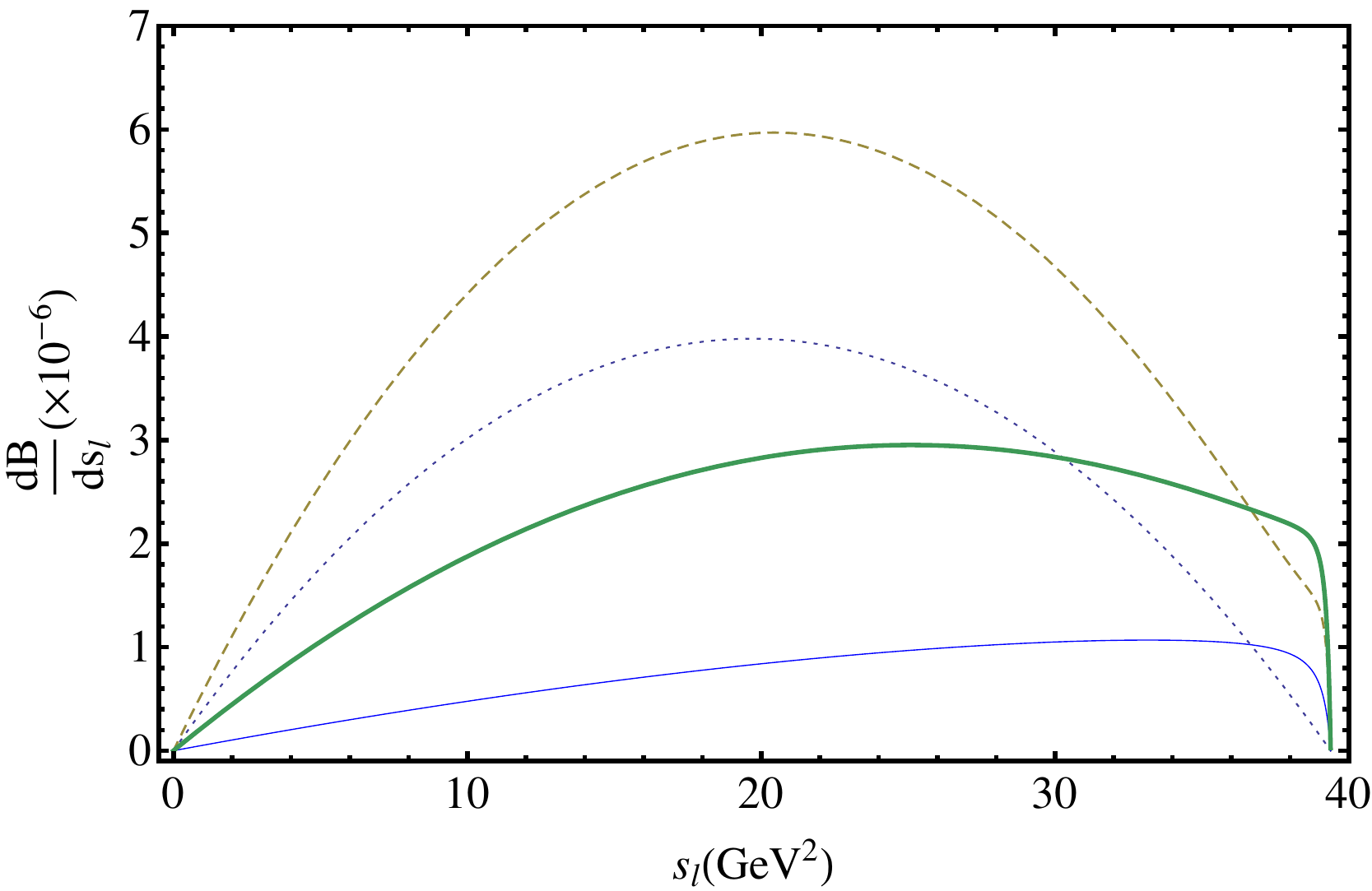}
\caption{ Similar with Fig.~\ref{fig:ek} but for the $s_l$ dependence.  }
\label{fig:s}
\end{figure}

\begin{table}[thb]
\caption{\label{tab:results}
Branching ratios of $B_c\to \gamma\ell\bar\nu$ and $B_c\to \ell\nu$. Here $\tau_{B_c}=0.50${ps},
and we vary the heavy quark masses with $m_b=4.8\pm0.1${GeV} and $m_c=1.5\mp0.1${GeV}.}
\begin{center}
\begin{tabular}{cccccc}
  \hline \hline
  Channels & Tree-level& $|\bold{v}|^2$-corrections& QCD corrections & This work (NLO)\\  \hline
  $B_c \to \tau \bar\nu_\tau$ &  $2.90\times 10^{-2}$& $-0.54 \times 10^{-2}$& $-0.56^{+0.03}_{-0.04}\times 10^{-2}$ & $1.80^{+0.03}_{-0.04}\times 10^{-2}$ \\
  $B_c \to \mu \bar\nu_\mu$ & $12.10\times 10^{-5}$ & $-2.25\times 10^{-5}$ & $-2.32^{+0.14}_{-0.16}\times 10^{-5}$ & $7.53^{+0.14}_{-0.16}\times 10^{-5}$ \\
  $B_c \to e \bar\nu_e$ & $2.82 \times 10^{-9}$& $-0.53\times 10^{-9}$ & $-0.54^{+0.03}_{-0.04}\times 10^{-9}$ & $1.75^{+0.03}_{-0.04}\times 10^{-9}$ \\
 $ B_c \to \gamma\mu \bar\nu_\mu $ & $10.49^{+2.27}_{-1.80}\times 10^{-5}$ & $5.46^{+1.35}_{-1.07}\times 10^{-5}$ & $-7.68_{+1.54}^{-1.97}\times 10^{-5}$ & $8.23^{+1.65}_{-1.33}\times 10^{-5} $\\
   \hline \hline
\end{tabular}
\end{center}
\end{table}

\begin{table}[thb]
\caption{\label{tab:results2}
Branching ratios of $B_c\to \gamma\ell\bar\nu$ and $B_c\to \ell\nu$ compared with other theories or models,
including Lattice QCD (LQCD), Light front model (LFM), Constituent quark model (CQM). Here $\tau_{B_c}=0.50${ps} is adopted. }
\begin{center}
\begin{tabular}{ccccccc}
  \hline \hline
   & This work& LQCD~\cite{McNeile:2012qf} & LFM~\cite{Lih:1999it} & CQM~\cite{Chang:1999gn}  & Ref.~\cite{Chen:2015csa} & Ref.~\cite{Chiladze:1998ny}\\  \hline
  $10^2 B(B_c \to \tau \bar\nu_\tau)$ &$1.80^{+0.03}_{-0.04}$ & $2.12$& $1.52$ & 1.44 &1.8&1.6\\
  $10^5 B(B_c \to \mu \bar\nu_\mu)$ &$7.53^{+0.14}_{-0.16}$ & $8.86$ & $6.09$ & 6.2&7.6& 5.7\\
  $10^9 B(B_c \to e \bar\nu_e)$ & $1.75^{+0.03}_{-0.04}$& $2.06$ & $1.41$& 1.47& 1.7& 1.5\\
 $ 10^5 B(B_c \to \gamma\mu \bar\nu_\mu)$ & $8.23^{+1.65}_{-1.33}$ & -- & $2.2(5)$&4.71 &--&4.78\\
   \hline \hline
\end{tabular}
\end{center}
\end{table}

With the estimated long-distance matrix elements,
results for   differential distributions are given in Figs.~\ref{fig:ek} and \ref{fig:s}, where the QCD and relativistic corrections are shown respectively.
The integrated  branching ratios of $B_c\to \gamma\ell\bar\nu$ and $B_c\to \ell\bar\nu$ are presented in Tabs.~\ref{tab:results} and \ref{tab:results2}. Ignoring the lepton mass, the branching ratio of $B_c\to \gamma e \bar\nu_e$ is identical to that of $B_c\to \gamma\mu \bar\nu_\mu$.
The LO results are in agreement with Ref.~\cite{Chang:1997re,Chiladze:1998ny,Lih:1999it,Colangelo:1999gb,Chang:1999gn} with  the same input parameters. From the calculation, one can see  that both the QCD
and relativistic corrections give   destructive  contributions to the process $B_c\to \ell\nu$. However, relativistic corrections  produce a constructive  contribution to the   $B_c\to \gamma\ell\bar\nu$. Our results have demonstrated that the QCD and relativistic corrections are mandatory towards a  more
accurate extraction of the value of LDMEs for $B_c$ system.

\section{Summary}
\label{sec:conclusions}

In this work, we have analyzed  the radiative leptonic $B_c\to \gamma\ell\bar\nu$ decays in the NRQCD effective field theory. NRQCD factorization ensures the separation of short-distance and long-distance effects of $B_c\to \gamma\ell\bar\nu$ into all order of $\alpha_s$. Treating the photon as a collinear object whose interactions with the heavy quarks can  be  integrated out, we arrive at a factorization formula for the decay amplitude.

We have calculated not only the    short-distance coefficients at leading order and next-to-leading order   in $\alpha_s$, but also the nonrelativistic corrections at the order  $|\bold{v}|^2$ in our analysis.  We found that the QCD corrections  can sizably  decrease the branching ratio,  which  has very important impact on  extracting the long-distance operator matrix elements of $B_c$. For phenomenological applications, we have estimated the long-distance matrix elements, which are further used to explore  the photon energy, lepton energy and lepton-neutrino invariant mass distribution. These results can be examined  at the  LHCb experiment.



\section*{Acknowledgments}
We are grateful to  Prof.  Yu Jia,  Cai-Dian L\"u  and Dr. Si-Hong Zhou for fruitful discussions. We thank the support of a key laboratory grant from the Office of Science and Technolny, Shanghai
Municipal Government (No. 11DZ2260700),  and  by Shanghai Natural  Science Foundation  under Grant No.15ZR1423100.

\begin{appendix}

\section{Ward identities for matrix elements}
\label{sec:ward_Iden}

In this section, we will derive the constraints on the $B_c\to \gamma$ form factors following  a Ward identity
for  the conservation of the electromagnetic current. To be more specific, let us
consider the
following matrix element:
\begin{eqnarray}\label{A1}
\langle \gamma(k,\epsilon)  |(\bar c\gamma_\nu\gamma_5 b)(0)|\overline B_c\rangle =
ie\epsilon^{*\mu}\int \mbox{d}^4x e^{ik\cdot x}
\langle 0|\mbox{T} j^{\rm e.m.}_\mu(x)\,(\bar c\gamma_\nu\gamma_5 b)(0)|\overline B_c\rangle\,.
\end{eqnarray}
In this case, the electromagnetic current includes contributions from
heavy quarks $j^{\rm e.m.}_\mu = e_c \bar c\gamma_\mu c + e_b\bar b\gamma_\mu
b$.

The conservation of the electromagnetic current implies  a Ward identity for the matrix element of the time-ordered
product in (\ref{A1})
\begin{eqnarray}
\label{A2}
&&ik^\mu \int \mbox{d}^4x e^{ik\cdot x}\langle 0|\mbox{T} j^{\rm e.m.}_\mu(x)\, |(\bar c\gamma_\nu\gamma_5 b)(0)|\overline B_c\rangle\nnbe
\int \mbox{d}^3x e^{ik\cdot x}(\langle 0|j^{\rm e.m.}_0(x)\,(\bar c\gamma_\nu\gamma_5 b)(0)|\overline B_c\rangle \theta(x^0)+\langle 0|(\bar c\gamma_\nu\gamma_5 b)(0)\,j^{\rm e.m.}_0(x)|\overline B_c\rangle \theta(-x^0))\big|^{x^0\to \infty}_{x^0\to -\infty} \nnbe
\int \mbox{d}^3x e^{-i\vec{k}\cdot \vec{x}}(\langle 0|j^{\rm e.m.}_0(\vec{x})\,(\bar c\gamma_\nu\gamma_5 b)(0)|\overline B_c\rangle -\langle 0|(\bar c\gamma_\nu\gamma_5 b)(0)\,j^{\rm e.m.}_0(\vec{x})|\overline B_c\rangle ) \nnbe
\int \mbox{d}^3x e^{-i\vec k\cdot \vec x}
\langle f|[j^{\rm e.m.}_0(\vec x)\,, (\bar c\gamma_\nu\gamma_5 b)(\vec 0)]|\overline B_c\rangle\,.
\end{eqnarray}
The commutator on the right-hand side is non-vanishing since the
operator $\bar c\gamma_\nu\gamma_5 b$ carries an electric charge. It can be evaluated as:
\begin{eqnarray}
\label{A3}
&&\int \mbox{d}^3x e^{-i\vec k\cdot \vec x}
\langle 0|[j^{\rm e.m.}_0(\vec x)\,, (\bar c\gamma_\nu\gamma_5 b)(\vec 0)] |\overline B(p_{B_c})\rangle\nnbe
\int \mbox{d}^3x e^{-i\vec k\cdot \vec x}
\langle 0|[e_c c_m^\dag(\vec{x}) c_m(\vec{x}) + e_b b_m^\dag(\vec{x}) b_m(\vec{x})\,,
c_n^\dag(0) (\gamma^0\gamma_\nu\gamma_5)_{ns} b_s(0)] |\overline B(p_{B_c})\rangle
\nnbe
 (e_c - e_b)
\langle 0|(\bar c\gamma_\nu\gamma_5 b)(\vec 0)|\overline B_c(p_{B_c}) \rangle\nonumber\\
&=& i(e_c - e_b) f_{B_c} p_{B_c,\nu}\,.
\end{eqnarray}
The most general parametrization of the matrix element on the left-hand side without $k^\mu$
can be written in terms of five form factors $f_i(k^2,p_{B_c}\cdot k)$
\begin{eqnarray}
i\int \mbox{d}^4x e^{ik\cdot x}
\langle 0|\mbox{T} j^{\rm e.m.}_\mu(x)\, (\bar
c\gamma_\nu\gamma_5 b)(0)|\overline B_c\rangle & =&
i[f_1 g_{\mu\nu} + f_2 p_{B_c,\mu} p_{B_c,\nu} + f_3 k_\mu k_\nu \nonumber\\&&+
f_4 k_\mu p_{B_c,\nu}+ f_5 p_{B_c,\mu} k_\nu]\,.
\end{eqnarray}
The Ward identity (\ref{A3}) implies two constraints on these
form factors
\begin{eqnarray}
(p_{B_c}\cdot k) f_2 + k^2 f_4 = (e_c-e_b) f_{B_c}\,,\qquad
f_1 + k^2 f_3 + (p_{B_c}\cdot k) f_5 = 0\,.
\end{eqnarray}
For  a real photon $k^2=0$, these constraints
fix uniquely the form factor $f_2(0,p_{B_c}\cdot k)$, and relate
$f_1(0,p_{B_c}\cdot k)$ and $f_5(0,p_{B_c}\cdot k)$, which leads to
\begin{eqnarray}
 \langle \gamma(\epsilon, k)|\bar c\gamma_\mu\gamma_5 b |\overline B_c (p_{B_c})\rangle &=& i e p_{B_c}\cdot k f_5\left(\epsilon_{\mu}^*  -k_\mu \frac{p_{B_c}\cdot \epsilon^*}{p_{B_c}\cdot k}  \right)-\frac{i e}{p_{B_c}\cdot k}f_{B_c} p_{B_c\mu} p_{B_c}\cdot \epsilon^* .~~~~~
\end{eqnarray}
This is the same as the result in Eq.~\eqref{eq:axial-vector-ff} as presented in text,
with the identification $p_{B_c}\cdot k f_5 = A$.

\section{Passarino-Veltman integrals\label{ap-Integral}}

The coefficients $b_i$, $c_i$ and $d_i$ are related to the scalar Passarino-Veltman integrals defined in Ref.~\cite{Passarino:1978jh,Hahn:1998yk}, and
we have split the finite pieces $b_i=B_i^{{finite}}$, $c_i=C_i^{{finite}}/m_b^2$ and $d_i=D_i^{{finite}}/m_b^4$:
\begin{eqnarray}
B_1&=&B_0\left(0,z^2m_b^2,z^2 m_b^2\right),\nonumber\\
B_2&=&B_0\left(0,m_b^2, m_b^2\right),\nonumber\\
B_3&=&B_0\left(m_b^2 \left(y^2-z \tilde{z}\right)/\tilde{z},0,m_b^2\right),\nonumber\\
B_4&=& B_0\left(y^2 m_b^2,m_b^2,z^2 m_b^2\right),\nonumber\\
C_1&=&\text{C}_0\left(m_b^2,0,m_b^2 \left(y^2-z \tilde{z}\right)/\tilde{z},0,m_b^2,m_b^2\right),\nonumber\\
C_2&=&\text{C}_0\left(\tilde{z}^2 m_b^2,y^2 m_b^2,0,m_b^2,z^2 m_b^2,m_b^2\right),\nonumber\\
C_3&=&\text{C}_0\left(m_b^2,z^2 m_b^2,\tilde{z}^2 m_b^2,m_b^2,0,z^2 m_b^2\right),\nonumber\\
C_4&=&\text{C}_0\left(m_b^2 \left(y^2-z \tilde{z}\right)/\tilde{z},m_b^2y^2, m_b^2z^2, 0, m_b^2, m_b^2z^2 \right),\nonumber\\
D_1&=&\text{D}_0\left(m_b^2,z^2 m_b^2,y^2 m_b^2,0,\tilde{z}^2 m_b^2,m_b^2 \left(y^2-z \tilde{z}\right)/\tilde{z},m_b^2,0,z^2 m_b^2,m_b^2\right).
\end{eqnarray}


Here  we give  the the results of divergence integrals.
\begin{eqnarray}
B_1&=&\frac{1}{\epsilon_{UV}}+\ln  \frac{\mu ^2}{z^2m_b^2},\nonumber
\\B_2&=&\frac{1}{\epsilon_{UV}}+\ln  \frac{\mu ^2}{m_b^2},\nonumber
\\B_3&=&\frac{1}{\epsilon_{UV}}+\ln  \frac{\mu ^2}{m_b^2} -\frac{ (y^2-\tilde{z}^2 ) \ln (\tilde{z}-\frac{y^2}{\tilde{z}} )}{y^2-z \tilde{z}}+2,\nonumber
\\B_4&=& \frac{1}{\epsilon_{UV}}+\ln  \frac{\mu ^2}{y^2m_b^2} +2+\sum_{i=1}^2(\gamma_i(y)\ln(\frac{\gamma_i(y)-1}{\gamma_i(y)})-\ln(\gamma_i(y)-1)),\nonumber
\\C_3&=& -\frac{1}{2zm_b^2}(\frac{1}{\epsilon_{IR}}+t_1+\ln \frac{\mu^2}{m_b^2}-2-\frac{2\ln z}{1+z}),\nonumber
\\D_1&=&\frac{\tilde{z}}{2 m_b^4 z \left(\tilde{z}^2-y^2\right)}(\frac{1}{\epsilon_{IR}}+t_1+\ln \frac{\mu^2}{m_b^2}-2\ln \frac{\tilde{z}^2-y^2}{\tilde{z}}+\frac{1}{(y-\tilde{z}) (y+\tilde{z})}(2  (\tilde{z}^2 \nonumber\\&&-2 y^2 \ln y- (y^2+z^2-1 ) \ln z-y^2 (1+2\ln 2) )+(-g_5+y^2+z^2-1)g_1\nonumber\\&&+(g_5+y^2-z^2+1)g_2+(-g_5+y^2-z^2+1)g_3+(g_5+y^2+z^2-1)g_4)),~~
\end{eqnarray}
where
\begin{eqnarray}
\gamma_{1,2}(x)&=&\frac{\pm\sqrt{\left(x^2-z^2+1\right)^2-4 x^2}+x^2-z^2+1}{2 x^2},\nonumber\\
g_1&=&\ln \left(\sqrt{\left(y^2-z^2+1\right)^2-4 y^2}-y^2-z^2+1\right),\nonumber\\
g_2&=& \ln \left(\sqrt{\left(y^2-z^2+1\right)^2-4 y^2}+y^2-z^2+1\right),\nonumber\\
g_3&=& \ln \left(-\sqrt{\left(y^2-z^2+1\right)^2-4 y^2}+y^2-z^2+1\right),\nonumber\\
g_4&=&\ln \left(-\sqrt{\left(y^2-z^2+1\right)^2-4y^2}-y^2-z^2+1\right),\nonumber\\
g_5&=&\sqrt{y^4-2 y^2 \left(z^2+1\right)+\left(z^2-1\right)^2}.
\end{eqnarray}

\section{One loop corrections to the axial-vector form factor $A$\label{ap-A}}

The most general structure of the matrix element of the axial-vector current is parametrized by:
\begin{eqnarray}
 \langle \gamma(\epsilon, k)|\bar c\gamma_\mu\gamma_5 b |[\bar cb ({}^1S_0^{[1]})]\rangle &=& i e\left(\epsilon_{\mu}^* \mathbb {A}^\epsilon -k_\mu \frac{p_{B_c}\cdot \epsilon^*}{p_{B_c}\cdot k}   \mathbb {A}^k\right)-i e\frac{ p_{B_c}\cdot \epsilon^*}{p_{B_c}\cdot k}\mathbb {f}^A p_{B_c\mu}. \label{eq:decay_A_One_Loop_NRQCD}
\end{eqnarray}
This section will be devoted to  demonstrate the gauge invariance at the one-loop level in $\alpha_s$, namely
\begin{eqnarray}
\mathbb {A}^\epsilon = \mathbb {A}^k\equiv \mathbb {A},\\
\mathbb {f}^A=\mathbb {f}.
\end{eqnarray}
The contributions from individual diagrams to $\mathbb {A}^\epsilon$ are given as
\begin{eqnarray}
\mathbb {A}^\epsilon_{b} &=& \frac{e_b C_F \alpha_s\sqrt{2N_c}}{4\pi m_b}\bigg[\frac{1}{\hat \epsilon_{IR}}-\frac{4 (z-1) \tilde{z}^2}{\left(y^2-\tilde{z}^2\right)^2}+\frac{y^2 \tilde{z}^2-2 (z-1) \tilde{z}^3-y^4}{\left(y^2-\tilde{z}^2\right)^2\tilde{z}}b_1+\frac{y^2}{(y^2-\tilde{z}^2)\tilde{z}}b_2\nonumber\\
&&+\frac{ 2 y^2 \tilde{z}-y^4+\left(z^2-1\right)^2 }{z \left(y^2-\tilde{z}^2\right)^2}b_3+\frac{ -2 y^2 \tilde{z}+(z-1) \tilde{z}^3+y^4 }{z \left(y^2-\tilde{z}^2\right)^2}b_4\nonumber\\
&&-\frac{2 (3 z-1) \tilde{z}^2+y^4-y^2 \left(z^2+4 z+3\right)}{  \left(\tilde{z}^2-y^2\right)\tilde{z}}c_4-\frac{\left(z^2-1\right) \left(y^2-z^2+1\right)}{(y^2-\tilde{z}^2)\tilde{z}}c_3\nonumber\\
&& + \left(-y^2+z^2+4 z-1\right)d_1\bigg], \\
\mathbb {A}^\epsilon_{c} &=&   \frac{e_b C_F \alpha_s\sqrt{2N_c}}{4\pi m_b}\bigg[\frac{1}{2}\frac{1}{\hat \epsilon_{UV}}+\frac{   \tilde{z}+y^2 }{y^2-\tilde{z}^2}b_1+\frac{   -y^2 (2 z+3)+2 z^3+5 z^2+2 z-1 }{2 z \left(\tilde{z}^2-y^2\right)}b_3\nonumber\\
&&+\frac{   y^2+z^2+4 z+3 }{y^2-\tilde{z}^2}+\frac{ \tilde{z} \left(-3 y^2+z^2-1\right)}{2 z \left(y^2-\tilde{z}^2\right)
  }b_4+  ({ -y^2+z^2+z-1})c_4\bigg], \\
\mathbb {A}^\epsilon_{d} &=&  \frac{e_b C_F \alpha_s\sqrt{2N_c}}{4\pi m_b}\bigg[\frac{1}{2}\frac{1}{\hat \epsilon_{UV}}+\frac{  y^2-z^2+1 }{y^2-\tilde{z}^2}+\frac{  y^2-z \tilde{z}}{y^2-\tilde{z}^2}b_2+-\frac{ y^2-z^2+1}{2 \left(y^2-\tilde{z}^2\right)}b_3-\tilde{z}c_1\bigg], \\
\mathbb {A}^\epsilon_{e} &=&   \frac{e_b C_F \alpha_s\sqrt{2N_c}}{4\pi m_b}\bigg[- \frac{1}{2\hat\epsilon_{UV}}+\frac{
y^2-z^2+1 }{2 \left(y^2-z \tilde{z}\right)}+\frac{\tilde{z}}{2 y^2-2 z \tilde{z}}b_2-\frac{ y^2-z^2+1}{2 \left(y^2-z \tilde{z}\right)}b_3\bigg],
\end{eqnarray}
The mass counter term and wave function renormalization give the contributions:
\begin{eqnarray}
\mathbb {A}^\epsilon_{CT-m} &=& 0,
\nonumber\\
\mathbb {A}^\epsilon_{CT-F} &=& -\mathbb {V}_{CT-F}.
\end{eqnarray}

The contributions from individual diagrams to $\mathbb {A}^k$ are given as
\begin{eqnarray}
\mathbb {A}^k_{b} &=&  \frac{e_b C_F \alpha_s\sqrt{2N_c}}{4\pi m_b}\bigg[-\frac{\tilde{z} \left(y^2 \left(-7 z^2+10 z+1\right) \tilde{z}^2+(z-1) \tilde{z}^5+y^4 \left(3 y^2+3 z^2-8 z-11\right)\right)}{\left(y^2-z \tilde{z}\right) \left(y^3-y \tilde{z}^2\right)^2}\nonumber\\
&&+\frac{2 y^2 (3-2 z) \tilde{z}^2+z \tilde{z}^4+y^4 (-(z+2))}{\left(y^3-y \tilde{z}^2\right)^2}b_1+\frac{\tilde{z}+y^2}{y^4-y^2 z \tilde{z}}b_2 -\frac{2 \tilde{z} \left(-y^2+z^2+3\right)}{  \left(y^2-\tilde{z}^2\right) }c_4\nonumber\\
&&+\frac{-y^2 \tilde{z}^2 \left(y^2+z^2-4 z+5\right)+\left(z^3-3 z^2+5 z+1\right) \tilde{z}^3+y^6}{z \left(-y^2+z^2+z\right) \left(y^2-\tilde{z}^2\right)^2}b_3\nonumber\\
&&+\frac{y^2 (3 z-5) \tilde{z}^3-(z-1) \tilde{z}^5+y^4 \left(y^2+z^2-1\right)}{z \left(y^3-y \tilde{z}^2\right)^2}b_4\bigg ]+\mathbb {A}^\epsilon_{b}, \\
\mathbb {A}^k_{c} &=&   \frac{e_b C_F \alpha_s\sqrt{2N_c}}{4\pi m_b}[\frac{\tilde{z}
   \left(y^2 \left(-7 z^2+10 z+1\right) \tilde{z}^2+(z-1) \tilde{z}^5+y^4 \left(3 y^2+3 z^2-8 z-11\right)\right)}{\left(y^2-z \tilde{z}\right) \left(y^3-y \tilde{z}^2\right)^2}\nonumber\\
   &&+\frac{ 2 y^2 (2 z-3) \tilde{z}^2-z \tilde{z}^4+y^4 (z+2)}{\left(y^3-y
   \tilde{z}^2\right)^2}b_1+\frac{ \tilde{z}+y^2}{y^4-y^2 z \tilde{z}}b_2-\frac{2 \tilde{z} \left(-y^2+z^2+3\right)}{  \left(y^2-\tilde{z}^2\right) }c_4\nonumber\\
   &&-\frac{-y^2 \tilde{z}^2 \left(y^2+z^2-4 z+5\right)+\left(z^3-3 z^2+5 z+1\right) \tilde{z}^3+y^6}{z \left(z \tilde{z}-y^2\right) \left(y^2-\tilde{z}^2\right)^2}b_3\nonumber\\
   &&-\frac{y^2 (3 z-5) \tilde{z}^3-(z-1) \tilde{z}^5+y^4
   \left(y^2+z^2-1\right)}{z \left(y^3-y \tilde{z}^2\right)^2}b_4]+\mathbb {A}^\epsilon_{c},
\end{eqnarray}
\begin{eqnarray}
\mathbb {A}^k_{d} &=&  \mathbb {A}^\epsilon_{d}+\ \frac{e_b C_F \alpha_s\sqrt{2N_c}}{4\pi m_b}[\frac{\tilde{z} \left(5 y^2-5 z^2-6
   z-1\right)}{ \left(y^2-\tilde{z}^2\right) \left(y^2-z \tilde{z}\right)}+\frac{\tilde{z} \left(3 y^2-3 z^2-4 z-1\right)}{ \left(y^2-\tilde{z}^2\right)\left(y^2-z \tilde{z}\right)}b_2\nonumber\\
   &&+\frac{\tilde{z} \left(-3 y^2+3 z^2+4
   z+1\right)}{ \left(y^2-\tilde{z}^2\right) \left(y^2-z \tilde{z}\right)}b_3], \\
\mathbb {A}^k_{e} &=&  \mathbb {A}^\epsilon_{e}+ \frac{e_b C_F \alpha_s\sqrt{2N_c}}{4\pi m_b}[\frac{3 \tilde{z}}{y^2-\tilde{z}^2}\frac{1}{\hat\epsilon_{UV}}+\frac{\tilde{z}}{-y^2+z^2+z}+\frac{\tilde{z}^2}{ \left(y^2-\tilde{z}^2\right) \left(y^2-z \tilde{z}\right)}b_2\nonumber\\
&&+\frac{\tilde{z} \left(3 y^2-3 z^2-4 z-1\right)}{ \left(y^2-\tilde{z}^2\right) \left(y^2-z\tilde{z}\right)}b_3].
\end{eqnarray}
Similar, the mass counter-terms and wave function renormalization corrections give:
\begin{eqnarray}
\mathbb {A}^k_{CT-m} &=&  \frac{e_b C_F \alpha_s\sqrt{2N_c}}{4\pi m_b}[\frac{3 \tilde{z} }{\tilde{z}^2-y^2}(\frac{1}{\hat\epsilon_{UV}}+\ln\frac{\mu^2}{m_b^2}+\frac{4 }{3})],
\nonumber\\
\mathbb {A}^k_{CT-F} &=&\mathbb {A}^\epsilon_{CT-F}.
\end{eqnarray}
Adding the above contributions, one may derive the  relation $\mathbb {A}^\epsilon = \mathbb {A}^k$, which is guaranteed by gauge invariance.  One can obtain  the one-loop results for $\mathbb {A}$ by  adding up the anti-symmetrical part with $e_b\to e_c$ and $m_b\leftrightarrow m_c$.

The contributions from individual diagrams to $\mathbb {f}^A$ are given as
\begin{eqnarray}
\mathbb {f}^A_{b} &=& \frac{e_b C_F \alpha_s\sqrt{2N_c}}{4\pi}[-\frac{2}{\hat \epsilon_{IR}}+\frac{y^2 \left(3 z^2-6 z-1\right) \tilde{z}^2-(z-1) \tilde{z}^5+y^4 \left(-3 y^2+z^2+8 z+7\right)}{y^2 \left(y^2-\tilde{z}^2\right) \left(y^2-z \tilde{z}\right)}\nonumber\\
&&+\frac{-3 y^2 (z-1) \tilde{z}^2+z \tilde{z}^4-y^4}{y^2 \tilde{z}  \left(y^2-\tilde{z}^2\right)}b_1+\frac{-\left(y^4+3\right) z+\left(y^2-1\right) z^3+\left(y^2-3\right) z^2-1}{y^2 \tilde{z} \left(y^2-z \tilde{z}\right)}b_2\nonumber\\
&&+\frac{4 y^2 \tilde{z}+\left(z^2-4 z-1\right) \tilde{z}^2-y^4}{z  \left(y^2-\tilde{z}^2\right)\left(y^2-z \tilde{z}\right)}b_3+\frac{\tilde{z} \left(-(z-1) \tilde{z}^3+y^4+2 y^2 \left(z^2-z-2\right)\right)}{y^2 z  \left(y^2-\tilde{z}^2\right)}b_4\nonumber\\&&-2\tilde z c_4+4zc_3], \\
\mathbb {f}^A_{c} &=&  \frac{e_b C_F \alpha_s\sqrt{2N_c}}{4\pi}[-\frac{1}{\hat \epsilon_{UV}}+\frac{y^2 \left(-5 z^2+4
   z+1\right) \tilde{z}^2+(z-1) \tilde{z}^5+y^4 \left(y^2+3 z^2-2 z-5\right)}{y^2  \left(y^2-\tilde{z}^2\right) \left(y^2-z \tilde{z}\right)} \nonumber\\
&&-\frac{\tilde{z} \left(z \tilde{z}^2+y^2 (2-3 z)\right)}{y^4-y^2
   \tilde{z}^2}b_1 +\frac{\tilde{z}^2}{y^4-y^2 z \tilde{z}}b_2+\frac{-4 y^2 \tilde{z}+\left(-z^2+4 z+1\right) \tilde{z}^2+y^4}{z
    \left(y^2-\tilde{z}^2\right) \left(y^2-z \tilde{z}\right)}b_3\nonumber\\
&&+\frac{\tilde{z} \left((z-1) \tilde{z}^3-y^2 \left(y^2+2 z^2-2
   z-4\right)\right)}{y^2 z  \left(y^2-\tilde{z}^2\right)}b_4+2 \tilde{z} c_4], \\
\mathbb {f}^A_{d} &=&  \frac{e_b C_F \alpha_s\sqrt{2N_c}}{4\pi}[-\frac{1}{\hat \epsilon_{UV}}+\frac{\left(z^2+10 z+1\right) \tilde{z}^2+y^4-2 y^2
   \left(z^2+6 z+5\right)}{ \left(y^2-\tilde{z}^2\right) \left(y^2-z \tilde{z}\right)} \nonumber\\
   &&+\frac{\tilde{z} \left(-5 y^2+5 z^2+6
   z+1\right)}{ \left(y^2-\tilde{z}^2\right)\left(y^2-z \tilde{z}\right)}b_2-\frac{\left(z^2+6 z+1\right) \tilde{z}^2+y^4-2 y^2 \left(z^2+4 z+3\right)}{ \left(y^2-\tilde{z}^2\right) \left(y^2-z
   \tilde{z}\right)}b_3], \\
\mathbb {f}^A_{e} &=&  \frac{e_b C_F \alpha_s\sqrt{2N_c}}{4\pi}[\frac{y^2-z^2-8z-7}{y^2-\tilde{z}^2}\frac{1}{\hat\epsilon_{UV}}+\frac{\tilde{z}^2-y^2}{y^2-z \tilde{z}}+\frac{\tilde{z} \left(-y^2+z^2-1\right)}{ \left(y^2-\tilde{z}^2\right)\left(y^2-z
   \tilde{z}\right)}b_2\nonumber\\
   &&\frac{\left(z^2+6 z+1\right) \tilde{z}^2+y^4-2 y^2 \left(z^2+4 z+3\right)}{ \left(y^2-\tilde{z}^2\right) \left(y^2-z \tilde{z}\right)}b_3],
\end{eqnarray}
\begin{eqnarray}
\mathbb {f}^A_{CT-m} &=& -\frac{2}{\tilde{z}}\mathbb {A}^\epsilon_{CT-m},
\nonumber\\
\mathbb {f}^A_{CT-F} &=&-\frac{2}{\tilde{z}}\mathbb {A}^\epsilon_{CT-F}.
\end{eqnarray}
The sum of them is
\begin{eqnarray}\mathbb {f}^A_{b-e+CT}&=&-\frac{3 e_b C_F \alpha _s\sqrt{2N_c} \left((z-1) \ln (z)-2 \tilde{z}+2/3t_1\right)}{4 \pi  \tilde{z}}.
\end{eqnarray}

We can get the one-loop result in Eq.~\ref{c0f1} after adding up the symmetrical part with $e_b\to e_c$ and $m_b\leftrightarrow m_c$.

\end{appendix}


\end{document}